\documentclass[aps,prd,onecolumn,superscriptaddress,preprintnumbers,floatfix,nofootinbib,longbibliography,eqsecnum]{revtex4-2}
\usepackage[german,english]{babel}
\usepackage{amsmath,amssymb}
\usepackage{hyperref}
\usepackage{titlesec}
\usepackage{bm}
\usepackage{bbm}
\usepackage{bbold}
\usepackage{ulem}
\usepackage{braket}
\usepackage{tablefootnote}
\usepackage{slashed}
\usepackage{multirow}
\usepackage{epsfig}
\usepackage{epstopdf}
\usepackage{diagbox}
\usepackage[mathlines]{lineno}
\usepackage[utf8]{inputenc}
\usepackage{comment}
\usepackage{graphicx}
\usepackage{subcaption}
\def\la{\lambda}     \def\dag{\dagger}
   \def\Oc{{\rm O}} \def\S{{\rm S}}
\def\bnabla{{\bm \nabla}}

\newcommand{\RN}[1]{%
  \textup{\uppercase\expandafter{\romannumeral#1}}%
}

\newcommand{\beq}{\begin{equation}}
\newcommand{\eeq}{\end{equation}}

\newcommand{\be}{\begin{equation}}
\newcommand{\ee}{\end{equation}}
\newcommand{\bea}{\begin{eqnarray}}
\newcommand{\eea}{\end{eqnarray}}

\makeatletter
\let\LN@align\align
\let\LN@endalign\endalign
\renewcommand{\align}{\linenomath\LN@align}
\renewcommand{\endalign}{\LN@endalign\endlinenomath}
\let\LN@gather\gather
\let\LN@endgather\endgather
\renewcommand{\gather}{\linenomath\LN@gather}
\renewcommand{\endgather}{\LN@endgather\endlinenomath}
\makeatother
\allowdisplaybreaks
\begin{document}

\preprint{TUM-EFT 164/21}
\title{Heavy hybrid decays to quarkonia}
\author{Nora Brambilla}
\email{nora.brambilla@tum.de}
\affiliation{Physik-Department, Technische Universit\"at M\"unchen, \\ James-Franck-Str.~1, 85748 Garching, Germany}
\affiliation{Institute for Advanced Study, Technische Universit\"at M\"unchen, \\ Lichtenbergstrasse 2 a, 85748 Garching, Germany}
\affiliation{Munich Data Science Institute, Technische Universit\"at M\"unchen, \\ Walther-von-Dyck-Strasse 10, 85748 Garching, Germany}
\author{Wai Kin Lai}
\email{wklai@m.scnu.edu.cn}
\affiliation{Guangdong Provincial Key Laboratory of Nuclear Science, Institute of Quantum Matter, South China Normal University, Guangzhou 510006, China}
\affiliation{Guangdong-Hong Kong Joint Laboratory of Quantum Matter,
Southern Nuclear Science Computing Center, South China Normal University, Guangzhou 510006, China}
\affiliation{Department of Physics and Astronomy, University of California, Los Angeles, California 90095, USA}
\author{Abhishek Mohapatra}
\email{abhishek.mohapatra@tum.de}
\affiliation{Physik-Department, Technische Universit\"at M\"unchen, \\ James-Franck-Str.~1, 85748 Garching, Germany}
\author{Antonio Vairo}
\email{antonio.vairo@tum.de}
\affiliation{Physik-Department, Technische Universit\"at M\"unchen, \\ James-Franck-Str.~1, 85748 Garching, Germany}

\begin{abstract}
The decay rates of the XYZ exotics discovered in the heavy 
quarkonium sector  are crucial observables for identifying the nature of these  states. 
Based on the framework of nonrelativistic effective field theories, we calculate the rates of  
semi-inclusive decays of heavy quarkonium hybrids into conventional heavy quarkonia.
We compute them at leading and subleading power in the inverse of the heavy-quark mass, extending and updating previous results.
We compare our predictions with experimental data of inclusive decay rates for candidates of heavy quarkonium hybrids. 
\end{abstract}


\maketitle 
\clearpage

\section{Introduction}
\label{sec:intro}
Hadrons, as bound states of quarks and gluons, have long been a major arena for testing our understanding of the strong interactions.
Traditionally, in the quark model~\cite{GellMann:1964nj,Zweig:1964CERN}, the hadrons were classified into mesons, which are bound states of a quark-antiquark pair, or baryons, which are bound states of three quarks.
The meson-baryon paradigm classifies successfully all hadrons discovered before 2003.
Besides the conventional hadrons, the quark model also predicted the possibilities of tetraquarks ($4$-quark states) and pentaquarks ($5$-quark states).
With the advent of quantum chromodynamics (QCD), the color degrees of freedom opened up even more possibilities such as hybrids,
which are hadrons with gluonic excitations, and glueballs, which are bound states of gluons. 
Hadrons that fall outside the meson-baryon paradigm are known as  exotic hadrons.
The so called XYZ states\footnote{These states have been termed XYZ in the discovery publications, without any special criterion, apart from $Y$ being used for exotics with vector quantum numbers $J^{PC}= 1^{--}$. Meanwhile the particle data group proposed a new naming scheme that extends the one used for ordinary quarkonia, in which the new names carry information on the $J^{PC}$ quantum numbers, see~\cite{Brambilla:2019esw} and the PDG~\cite{Workman:2022ynf} for more information. Since the situation is still in evolution, in this paper we use both naming schemes.}
are candidates for exotic hadrons in the heavy quarkonium sector, containing a heavy quark and antiquark pair.
They are exotic because either their masses do not fit the usual heavy quarkonium spectra,
or have unexpected decay modes if interpreted as conventional quarkonia, or have exotic quantum numbers such as the charged $Z_c$ and $Z_b$ states.
In 2003,  the  Belle experiment observed the first exotic state $X(3872)$ \cite{Belle:2003nnu}.
Since then, dozens of new XYZ states have been observed by different experimental groups:
B-factories (BaBar, Belle and CLEO), $\tau$-charm facilities (CLEO-c, BESIII), and also proton-(anti)proton colliders (CDF, D0, LHCb, ATLAS, CMS)
(see Refs.~\cite{Olsen:2014qna,Brambilla:2019esw} for details on experimental observations).

Many interpretations have been proposed for the nature of the XYZ states: quarkonium hybrids, compact tetraquarks, diquark-diquarks, heavy meson molecules, and hadroquarkonia
(see e.g. Refs.~\cite{Olsen:2014qna,Brambilla:2019esw,Ali:2019roi} for some comprehensive reviews). However, no single interpretation can explain the entire spectrum of the XYZ states.
Since some of the new exotic states were discovered from their decays into traditional quarkonia, 
theoretical studies of these decay modes could potentially provide a mean to unveil the nature of the XYZ states. 

For heavy hybrids, an effective field theory description called Born--Oppenheimer effective field theory (BOEFT) has been proposed \cite{Berwein:2015vca,Oncala:2017hop,Brambilla:2017uyf,Soto:2020xpm}.
A heavy hybrid state consists of a heavy-quark-antiquark pair in a color octet configuration 
bound with gluons. 
The nonrelativistic motion of the heavy quark and antiquark evolves with a time scale that is much larger than the typical time scale of the nonperturbative gluon dynamics, $1/\Lambda_{\rm QCD}$.
This leads to a scenario that resembles the Born--Oppenheimer approximation in diatomic
molecules~\cite{Griffiths:1983ah,Juge:1997nc,Braaten:2013boa,Braaten:2014ita,Braaten:2014qka,Meyer:2015eta}. 
BOEFT takes advantage of this scale separation to construct a systematic description of the heavy hybrid multiplet spectra \cite{Berwein:2015vca}
to be compared eventually with the masses and the quantum numbers of the observed neutral exotic states.
Effects of the spin have been introduced in BOEFT through spin-dependent potentials
finding a contribution already at order $1/m_Q$~\cite{Soto:2017one,Oncala:2017hop,Brambilla:2018pyn,Brambilla:2019jfi},
$m_Q$ being the heavy-quark mass, 
which is at variance with the spin structure of the potential in heavy quarkonium, where spin dependent effects start at order $1/m_Q^2$.
This hints to a possible stronger breaking of the heavy-quark spin symmetry in observables like  spin multiplets and decays.

In this work, we use BOEFT  to compute hybrid decay rates.
In particular, our objective is to study the inclusive transition rate of a heavy quarkonium hybrid $H_m$ to a quarkonium $Q_n$,  i.e,\ $H_m\rightarrow Q_n + X$,
where $X$ denotes any final state made of light particles, under the assumption that the energy gap between the hybrid and the quarkonium state is much larger than $\Lambda_{\mathrm {QCD}}$.
Some of these decays have been addressed in  Refs.~\cite{Oncala:2017hop,TarrusCastella:2021pld}.
We adopt a similar approach, emphasizing the various  assumptions entering the computation, and extending and updating the analysis to states that respect
the hierarchy of energy scales that lies at the basis of the whole effective field theory (EFT) construction.
We obtain decay rates in the charmonium and bottomonium hybrid sector and compare with existing experimental data.
As we only calculate decays to quarkonium, our results provide lower bounds for the total widths of heavy hybrids.

The paper is organized as follows.
In Sec.~\ref{sec:spectrum}, we fix the quarkonium potential on the quarkonium energy levels and the hybrid potentials on lattice QCD data.
We also write for hybrid states the coupled Schr\"odinger equations that follow from BOEFT. 
In Sec.~\ref{sec:decay_rate}, we compute the imaginary parts of the hybrid potentials, and the hybrid-to-quarkonium decay rates.
In Sec.~\ref{sec:results}, we present an updated comparison of the obtained hybrid multiplets with experimental candidates, we report our results for the hybrid-to-quarkonium decay rates and compare with experimental data.
In Sec.~\ref{conclusion}, we summarize and conclude.

\section{Spectra}
\label{sec:spectrum}
The  QCD static energies associated to a $Q \bar{Q} $ pair (quarkonium) and to a $Q \bar{Q}$ pair 
bound to gluons (hybrid)
can be classified according to the quantum numbers of the cylindrical symmetry group $D_{\infty h}$  
similarly to what happens for a diatomic molecule in QED.
A remarkable feature is that in the short-distance region the static energies can be organized in quasi-degenerate multiplets corresponding to the
gluelump spectrum that bears the spherical symmetry $O\left(3\right)\otimes C$ \cite{Foster:1998wu,Brambilla:1999xf}.

The static energies are nonperturbative matrix elements defined by some generalized Wilson loops,
which have been calculated on the lattice for the case of the pure SU(3) gauge theory~\cite{Juge:1997nc, Bali:2003jq, Juge:2002br, Schlosser:2021wnr, Capitani:2018rox}.
A tower of states can be associated to each of these energies by solving the corresponding Schr\"{o}dinger equation(s).
In what follows, concerning the hybrids, we focus only on the two lowest static energies $\Pi_u$ and $\Sigma_u^-$ that are degenerate at short distance.
We ignore mixing with states built out of higher static energies; 
these are separated by a gap of at least 400~MeV, which is of the order of $\Lambda_{\mathrm {QCD}}$, 
from the static energies $\Pi_u$ and $\Sigma_u^-$.
 
The relevant energy scales to describe quarkonium and hybrid states made of heavy (nonrelativistic) quarks 
are the scale $m_Q$, the scale $m_Q v$, which is the typical momentum transfer between the heavy quarks, $v \ll 1$ being the velocity of the heavy quark in the bound state, 
the scale $m_Q v^2$, which is the typical heavy-quark-antiquark binding energy, and $\Lambda_{\mathrm {QCD}}$, which is the scale of nonperturbative physics.
Such scales are hierarchically ordered, $m_Q \gg m_Qv \gg m_Qv^2$, 
and allow to introduce a hierarchy of  nonrelativistic effective field theories~\cite{Brambilla:2004jw} that turn out to be instrumental to computing observables.

Nonrelativistic QCD (NRQCD) follows from QCD by integrating out modes associated with the bottom or the charm quark mass~\mbox{\cite{Caswell:1985ui, Bodwin:1994jh,Manohar:1997qy}}.
For quarkonium, integrating out the scale of the momentum transfer $m_Qv$ leads
either to weakly-coupled potential NRQCD (pNRQCD)~\cite{Brambilla:1999xf, Pineda:1997bj} if $m_Q v \gg \Lambda_{\mathrm {QCD}}$,
or strongly-coupled pNRQCD ~\cite{Brambilla:1999xf,Brambilla:2000gk,Pineda:2000sz} if $m_Q v \sim \Lambda_{\mathrm {QCD}}$.
Contributions from gluons of energy and momentum of order $m_Qv$ are encoded in the pNRQCD potentials.
Hybrids are rather extended objects.
For this reason we assume that gluons responsible for their binding satisfy the strongly-coupled hierarchy  $m_Q v \gtrsim \Lambda_{\mathrm {QCD}}$.
The assumption also guarantees that  $m_Q v^2 \ll \Lambda_{\mathrm {QCD}}$, which prevents,  
at least parametrically,
mixing between hybrid states separated by a gap of order $\Lambda_{\mathrm {QCD}}$
and enables the Born--Oppenheimer approximation to work.
We look at hybrid states that are excitations of the lowest lying static energies $\Pi_u$ and $\Sigma_u^-$. 
We need to consider both because they are degenerate in the short distance limit, which breaks the condition  $m_Q v^2 \ll \Lambda_{\mathrm {QCD}}$ for the associated hybrid states.
The hybrid states associated to the static energies $\Pi_u$ and $\Sigma_u^-$ mix
and the corresponding equations of motion are a set of coupled  Schr\"odinger equations~\cite{Berwein:2015vca}.
Higher static energies are separated by a gap of order $\Lambda_{\mathrm {QCD}}$ from the static energies  $\Pi_u$ and $\Sigma_u^-$,
and their modes are integrated out when integrating out gluons of energy or momentum of order $\Lambda_{\mathrm {QCD}}$.
Integrating out gluons of energy or momentum of order $\Lambda_{\mathrm {QCD}}$ or larger from NRQCD and
keeping quarkonium and hybrid states associated to the static energies $\Pi_u$ and $\Sigma_u^-$ as degrees of freedom
leads to an EFT that may be understood as an extension of strongly coupled pNRQCD to quarkonium hybrids.
This EFT is BOEFT, whose Lagrangian reads~\cite{Berwein:2015vca,Oncala:2017hop,Brambilla:2017uyf,Soto:2020xpm}:
\begin{equation}
L_{\mathrm{BOEFT}} = L_{\Psi}+L_{{\Psi}_{\kappa\lambda}}+L_{\mathrm{mixing}},
\label{eq:LBOEFT}\\
\end{equation}
with 
\begin{align}
&L_{\Psi}=
\int d^3{\bm R}\int d^3{\bm r} \; \mathrm{Tr}\Bigg[\Psi^{\dagger}(\mathbf{r},\,\mathbf{R},\,t)\left(i\partial_t+\frac{\bm{\nabla}^2_r}{m_Q}-V_{\Psi}(r)\right)\Psi(\mathbf{r},\,\mathbf{R},\,t)\Bigg],
\label{eq:LQQ}\\
&L_{{\Psi}_{\kappa\lambda}} = \int d^3{\bm R}\int d^3{\bm r} \, \sum_{\kappa\lambda\lambda^{\prime}}\mathrm{Tr}\left\{{\Psi}^{\dagger}_{\kappa\lambda}(\mathbf{r},\,\mathbf{R},\,t) \left[i\partial_t
                                  - V_{\kappa\lambda\lambda^{\prime}}(r)+ P^{i\dag}_{\kappa\lambda}\frac{\bm{\nabla}^2_r}{m_Q}P^i_{\kappa\lambda^{\prime}}
                   \right]{\Psi}_{\kappa\lambda^{\prime}}(\mathbf{r},\,\mathbf{R},\,t)\right\}, 
\label{eq:LQQg}\\
&L_{\mathrm{mixing}} = -\int d^3{\bm R}\int d^3{\bm r}\,\sum_{\kappa\lambda}\,\mathrm{Tr}\left[\Psi^{\dagger}\,V_{\kappa\lambda}^{\mathrm{mix}}\,{\Psi}_{\kappa\lambda} + \mathrm{h.c.}\right]\,,
    \label{eq:L_mixing}
\end{align}
where the trace is over the spin indices.
The fields $\Psi$ and ${\Psi}_{\kappa\lambda}$ denote the quarkonium and the hybrid fields, respectively.
They are functions of the relative coordinate $\bm{r}\equiv {\bm x}_1-{\bm x}_2$, and the center of mass coordinate $\bm{R}\equiv \left({\bm x}_1+{\bm x}_2\right)/2$ of the $Q\bar{Q}$ pair,
${\bm x}_1$ and ${\bm x}_2$ being the space locations of the quark and antiquark.
The label $\kappa = K^{PC}$ (${\mathbf K}$ is the angular momentum)  denotes the quantum numbers of the light degrees of freedom (LDF).
The projection vectors $P^i_{\kappa\lambda}$ ($i$ is the vector or spin index) project to an eigenstate of $\bm{K}\cdot\hat{\bm{r}}$
with eigenvalue $\lambda=-K, \dots, 0, \dots, K$, fixing the $D_{\infty h}$ quantum numbers.
The quarkonium and the hybrid potentials denoted in Eqs.~\eqref{eq:LQQ} and \eqref{eq:LQQg} by $V_{\Psi}$ and $V_{\kappa\lambda\lambda^{\prime}}$, respectively,
can be organized as expansions in $1/m_Q$,  
\begin{align}
&V_{\Psi} = V_{\Psi}^{(0)}(r)+\frac{V_{\Psi}^{(1)}(r)}{m_Q}+\dots,\label{eq:VQQbar}\\
&V_{\kappa\lambda\lambda^{\prime}}(r) \equiv
    P^{i\dagger}_{\kappa\lambda} V^{ij}_{\kappa}(r) P^j_{\kappa\lambda^{\prime}}=V^{(0)}_{\kappa\lambda}(r)\delta_{\lambda\lambda^{\prime}}+\frac{V^{(1)}_{\kappa\lambda\lambda^{\prime}}(r)}{m_Q}+\dots\,, 
\label{eq:VQQbarg}
\end{align}
where $V_{\Psi}^{(0)}(r)$ and $V^{(0)}_{\kappa\lambda}(r)$ are the quarkonium and the hybrid static potentials.
They are independent of the heavy quark spins and may be identified with the static energies computed in lattice QCD: 
$V_{\Psi}^{(0)}(r)=E_{\Sigma_g^+}\left(r\right)$, $V_{10}^{(0)}(r)=E_{\Sigma_u^-}\left(r\right)$, and
$V_{1\pm 1}^{(0)}(r)=E_{\Pi_u}\left(r\right)$.
The effective potentials may also contain imaginary parts accounting for the quarkonium and hybrid decay and transitions.
The imaginary parts affecting $V_{\kappa\lambda\lambda^{\prime}}$ and coming from hybrid to quarkonium transitions will be determined in section~\ref{sec:decay_rate}.
The hybrid-quarkonium mixing potential $V_{\kappa\lambda}^{\mathrm {mix}}$ in Eq.~\eqref{eq:L_mixing} is of order $1/m_Q$ and depends on the spin of the heavy quark and the heavy antiquark.
In the current work, we ignore the effect from mixing, and therefore set $V_{\kappa\lambda}^{\mathrm {mix}}=0$ (see Ref.~\cite{Oncala:2017hop} for details on the mixing).

For quarkonium, the quantum numbers of the LDF are $\kappa=0^{++}$, which implies a trivial form of the projection vector, $P_{00}=\mathbb{1}$.
For low-lying hybrid states that are excitations of the static energies  $\Pi_u$ and $\Sigma_u^-$, the quantum numbers of the LDF are $\kappa=1^{+-}$.
The projection vectors $P^{i}_{\kappa\lambda}\equiv P^{i}_{1\lambda}$ read 
\begin{equation}
    P^i_{10} = \hat{r}^i,\qquad
    P^i_{1\pm 1}=\hat{r}_{\pm}^{i}\equiv\left(\hat{\theta}^i\pm i\hat{\phi}^i\right)/\sqrt{2},\label{eq:P_1}
\end{equation}
where $\hat{r}$, $\hat{\theta}$, and $\hat{\phi}$ are the spherical unit vectors.

In the following Secs.~\ref{subsec:Quarkonium} and~\ref{subsec:hybrids}, we compute the quarkonium and hybrid spectra from the static quarkonium and hybrid potentials.
This allows us to determine the quarkonium and hybrid wavefunctions and masses.
They will be necessary in Sec.~\ref{sec:decay_rate} where we compute the imaginary part of the hybrid potential coming from hybrid to quarkonium transitions and
the corresponding transition rates.

\subsection{Quarkonium}\label{subsec:Quarkonium}
Quarkonium states $(Q\bar{Q})$  are  color singlet bound states of a $Q\bar{Q}$ pair with static potential $E_{\Sigma_g^+}(r)$.
The leading order equation of motion for the field $\Psi({\bm r}, {\bm R}=0)$ that follows from Eq.~\eqref{eq:LQQ} is the Schr\"{o}dinger equation:
\begin{equation}
	\label{eq:Sch_QQbar}
	\left(-\frac{\bm{\nabla}^2}{m_Q}+E_{\Sigma_g^+}(r)\right)\Phi^{Q\bar{Q}}_{(n)}({\bm r})=E_n^{Q\bar{Q}}\, \Phi^{Q\bar{Q}}_{(n)}({\bm r})\,,
\end{equation}
where $E_n^{Q\bar{Q}}$ is the quarkonium energy and $\Phi^{Q\bar{Q}}_{(n)}\left({\bm r}\right)$ denotes the quarkonium wavefunction, which is related to the field operator $\Psi(\bm{r},{\bm R}={\bm 0})$ by
\begin{align}
\Phi^{Q\bar{Q}}_{(n)}\left({\bm r}\right)&=\langle 0|\Psi(\bm{r},  {\bm R}={\bm 0})|Q_n\rangle\,,
\label{eq:QQPsiwf}
\end{align} 
$|Q_n\rangle$ being the quarkonium state with quantum numbers $ n \equiv \{n, j, m_j, l, s\}$.
Including the spin and angular dependence, the complete quarkonium wavefunction is given by
\begin{equation}
\Phi^{Q\bar{Q}}_{(n)}({\bm r})\equiv\Phi^{Q\bar{Q}}_{(n, j, m_j,  l, s)}({\bm r})=\sum_{m_l,m_s}{\cal C}_{l m_l s m_s}^{j m_j}\frac{R_{nl}(r)}{r}\,Y_{l}^{m_l}\left(\theta, \phi\right)\chi_{s m_s}\,.
\label{eq:QQbarwf}
\end{equation}
If we call ${\bm L_{Q\bar{Q}}}$ the $Q\bar{Q}$ pair orbital angular momentum, ${\bm S}={\bm S_1}+{\bm S_2}$ the $Q\bar{Q}$ pair total spin, and ${\bm J}={\bm L_{Q\bar{Q}}}+{\bm S}$ the total angular momentum,
the quantum numbers are as follows: $n$ is the principal quantum number, $l(l+1)$ is the eigenvalue of $\bm{L}_{Q\bar{Q}}^2$, $j(j+1)$ and $m_j$ are the eigenvalues of $\bm{J}^2$ and $J_3$, respectively,
and $s(s+1)$ is  the eigenvalue of $\bm{S}^2$.
The functions $\chi_{s m_s}$ are the spin wavefunctions and
$\mathcal{C}^{j m_j}_{l m_l s m_s}$
are suitable Clebsch--Gordan coefficients.
The functions $R_{nl}(r)/r$ are the radial wavefunctions. 

The shape of the static potential $E_{\Sigma_g^+}(r)$ computed in lattice QCD is well described by a Cornell potential:
\begin{align}
E_{\Sigma_g^+}(r)= -\frac{\kappa_g}{r}+\sigma_g r + E_g^{Q}.
\label{eq:VSigmag}
\end{align}
The parameters $\kappa_g$ and the string tension $\sigma_g$ fitted to the lattice data give~\cite{Juge:2002br}:
\begin{equation}
\kappa_g=0.489 \, ,\hspace{1cm} \sigma_g=0.187 \,\mathrm{GeV}^2 \,.
\end{equation} 
For computing the quarkonium spectrum, we use the renormalon subtracted (RS) charm and bottom masses~\cite{Pineda:2001zq,Bali:2003jq}
defined at the renormalon subtraction scale $\nu_f=1\,\mathrm{GeV}$: $m_c^{\mathrm{RS}}= 1.477\,\mathrm{GeV}$ and $m_b^{\mathrm{RS}}=4.863\,\mathrm{GeV}$, which are the values used in  Ref.~\cite{Berwein:2015vca}.
Following Ref.~\cite{Oncala:2017hop}, once the quark masses have been assigned, the values of the offset $E_g^{Q}$ in Eq.~\eqref{eq:VSigmag} are tuned separately for charmonium and bottomonium states 
to best agree with the experimental spin-average masses~\cite{Workman:2022ynf}:
\begin{equation}
E_g^{c} = -0.254\,\mathrm{GeV},\qquad E_g^{b}= -0.195\,\mathrm{GeV}.
\label{eq:E_gQQbar}
\end{equation}
The numerical solutions of the Schr\"odinger equation \eqref{eq:Sch_QQbar} for some $S$-wave 
and $P$-wave charmonia and bottomonia 
below threshold 
are shown in Table~\ref{tab:QQbarspectrum}.
The corresponding radial wavefunctions $R_{nl}(r)$ are shown in appendix~\ref{app:QQbarwf}.

\begin{table}[h!]
\begin{center}
  \begin{tabular}{|c||c|c||c|c|}    
		\hline                                                                     
		$nL$  &  $M_{c\bar{c}}$    & $E_{\mathrm{exp}}$& $M_{b\bar{b}}$    & $E_{\mathrm{exp}}$ \\
		\hline \hline
$1S$ & 3068 & 3068 & 9442 & 9445\\
$2S$ & 3678  & 3674 & 10009 & 10017\\ 
$3S$ &     &  & 10356  & 10355\\ 
\hline
$1P$ & 3494  & 3525 & 9908  & 9900 \\ 
$2P$ &    &   & 10265  & 10260  \\ 
$3P$ &    &   & 10554 &  \\ 
\hline
	\end{tabular}
\caption{Spin-averaged masses (in MeV) of $S$- and $P$-wave charmonium and bottomonium states below threshold computed using the static potential in Eq.~\eqref{eq:VSigmag}.
The quarkonium mass is given by $M_{Q\bar{Q}}=2m_Q + E_n^{Q\bar{Q}}$ with  $Q=c$, $b$.
We use the charm and bottom masses: $m_c^{\mathrm{RS}}= 1.477\,\mathrm{GeV}$ and $m_b^{\mathrm{RS}}=4.863\,\mathrm{GeV}$.
$E_{\mathrm{exp}}$ denotes the spin-averaged experimental masses~\cite{Workman:2022ynf}.
We show only states relevant for this work.}
\label{tab:QQbarspectrum}
\end{center}
\end{table}

\subsection{Hybrids}\label{subsec:hybrids}
Hybrids $(Q\bar{Q}g)$ are exotic hadrons that are color-singlet bound states of a color octet $Q\bar{Q}$ pair coupled to gluons.
We focus here on the lowest-lying hybrid states that can be built from the $\Sigma_u^{-}$ and $\Pi_u$ static energies corresponding to LDF with quantum numbers $\kappa=1^{+-}$; 
from now on we drop the subscript $\kappa=1^{+\,-}$ if not necessary.
For $\kappa=1^{+-}$ three values of $\lambda$ (0 and $\pm 1$) are possible; for each value of $\lambda$ we can define a wavefunction in terms of the 
field operator ${\Psi}_\lambda(\bm{r},{\bm R}={\bm 0})$ acting on a hybrid state $|H_m\rangle$ with quantum numbers $ m \equiv \{m, j, m_j, l, s\}$:
\begin{equation}
\Psi^{(m)}_\lambda(\bm{r}) =\langle 0|{\Psi}_\lambda(\bm{r},\bm{R}=0)|H_m\rangle\,.
\label{eq:Psiwf}
\end{equation}
Hence, we can write in the hybrid rest frame
\begin{equation}
|H_m\rangle = \int d^3\bm{r}  
\sum_\lambda \Psi^{(m)}_\lambda(\bm{r}) \, {\Psi}^\dagger_\lambda(\bm{r},\bm{R}=0)|0\rangle\,.
\label{eq:Psiwf2}
\end{equation}
The quantum numbers are defined in the following way:
$m$ is the principle quantum number,
$l(l+1)$ is the eigenvalue of $\bm{L}^2$, $\bm{L}=\bm{L}_{Q\bar{Q}}+\bm{K}$ being the orbital angular momentum
sum of the orbital angular momentum of the $Q\bar{Q}$ pair and the angular momentum of the gluelump,
$s(s+1)$ is the eigenvalue of $\bm{S}^2$, ${\bm S}={\bm S_1}+{\bm S_2}$ being the spin of the $Q\bar{Q}$ pair, 
and $j(j+1)$ and $m_j$ are the eigenvalues of $\bm{J}^2$ and $J_3$ respectively, $\bm{J}=\bm{L}+\bm{S}$ being the total angular momentum. 

The wavefunctions $\Psi^{(m)}_\lambda$ are eigenfunctions of $\bm{K}\cdot\hat{\bm{r}}$ but not of parity.
The eigenfunctions of parity are called $\Psi_\Sigma^{(m)}$ and $\Psi_{\pm\Pi}^{(m)}$ and are linear combinations of $\Psi^{(m)}_\lambda(\bm{r})$.
The wavefunction $\Psi_\Sigma^{(m)}$  transforms as the spherical harmonics under parity, whereas  $\Psi_{\mp\Pi}^{(m)}$ transform with the same or with the opposite parity of $\Psi_\Sigma^{(m)}$. 
The parity eigenfunctions can be written as\footnote{
  Recall that $\bm{\hat{r}}_{\pm}$ and  $\bm{\hat{r}}$ project on states with definite  $\bm{K}\cdot\hat{\bm{r}}$, see Eq.~\eqref{eq:P_1}.
Hence, $\Psi_{\Sigma}^{(m)}$ is also an eigenfunction of $\bm{K}\cdot\hat{\bm{r}}$ with eigenvalue $\lambda = 0$.} 
\begin{align}
&  \Psi_{\Sigma}^{(m)}\left({\bm r}\right) =
  \sum_{m_l,\,m_s}\,{\cal C}_{j m_j l s}^{m_l m_s} v_{l,\,m_l}^0\,\bm{\hat{r}}\,\psi^{(m)}_\Sigma(r)\,\chi_{s m_s},
                \label{eq:Hybwf-1}\\
& \Psi_{\pm\Pi}^{(m)}\left({\bm r}\right) =
  \sum_{m_l,\,m_s}\frac{{\cal C}_{j m_j l s}^{m_l m_s}}{\sqrt{2}}\left(v_{l,\,m_l}^1\,\bm{\hat{r}}_{+}\pm v_{l,\,m_l}^{-1}\,\bm{\hat{r}}_{-}\right)\psi^{(m)}_{\pm\Pi}(r)\,\chi_{s m_s},
 \label{eq:Hybwf-2}
\end{align}
where the functions $\chi_{s m_s}$ are the spin wavefunctions and $\mathcal{C}_{j m_j l s}^{m_l m_s}$ are suitable Clebsch--Gordan coefficients.
The angular eigenfunctions $v_{l,\,m_l}^{\lambda}$ are generalizations of the spherical harmonics for systems with cylindrical symmetry~\cite{Landau:1991wop}.
Note that the hybrid wavefunctions in Eqs.~\eqref{eq:Hybwf-1} and \eqref{eq:Hybwf-2} are vector wavefunctions.
The functions $\psi^{(m)}_\Sigma$ and $\psi^{(m)}_{\pm\Pi}$ are radial wavefunctions.
Their equations may be derived from the equations of motion of the BOEFT Lagrangian \eqref{eq:LQQg}.
Since the  static energies $E_{\Sigma_{u}^{-}}$ and $E_{\Pi_u}$ mix in the short distance, the equations are a set of coupled Schr\"odinger equations.
Ignoring all corrections to the potentials but the static energies $E_{\Sigma_{u}^{-}}$ and $E_{\Pi_u}$, they read~\cite{Berwein:2015vca}:
\begin{align}
  \hspace{0.4 pt}&\left[-\frac{1}{m_Qr^2}\,\partial_rr^2\partial_r+\frac{1}{m_Qr^2}\begin{pmatrix} l(l+1)+2 & 2\sqrt{l(l+1)} \\
      2\sqrt{l(l+1)} & l(l+1) \end{pmatrix}+\begin{pmatrix} E_{\Sigma_{u}^{-}} & 0 \\
      0 & E_{\Pi_{u}} \end{pmatrix}\right]\hspace{-4pt}\begin{pmatrix} \psi_\Sigma^{(m)} \\
     \psi_{-\Pi}^{(m)}\end{pmatrix} = E_m^{Q\bar{Q}g} \begin{pmatrix} \psi_\Sigma^{(m)} \\
    \psi_{-\Pi}^{(m)}\end{pmatrix}\,,\nonumber\\
 &\hspace{4.0 cm}\left[-\frac{1}{m_Qr^2}\,\partial_r\,r^2\,\partial_r+\frac{l(l+1)}{m_Qr^2}+E_{\Pi_{u}}\right]\psi_{+\Pi}^{(m)} = E_m^{Q\bar{Q}g} \,\psi_{+\Pi}^{(m)}\,,
\label{eq:diffeq}
\end{align}
where $E_m^{Q\bar{Q}g}$ is the hybrid energy.

The set of Schr\"odinger equations \eqref{eq:diffeq} has no spin-dependence, so,  all the different spin configurations appear as degenerate multiplets.
The $J^{PC}$ quantum numbers are  $\left\{l^{\pm\pm};(l-1)^{\pm\mp},l^{\pm\mp},(l+1)^{\pm\mp}\right\}$,
where the first entry corresponds to the spin-$0$ combination and the next three entries to the spin-$1$ combinations.
For $l=0$, there is only one spin-$1$ combination as well as only one parity or charge conjugation state.
In Table~\ref{multiplets}, we show the first five degenerate multiplets.
The wavefunctions $\Psi^{(m)}_{\Sigma, -\Pi}({\bm r})$ describe the hybrid multiplets $H_1$, $H_3$, and $H_4$, 
while the  wavefunction $\Psi^{(m)}_{+\Pi}({\bm r})$ describes the hybrid multiplets $H_2$ and $H_5$. 

\begin{table}[t]
\begin{tabular}{|c|c|c|c|c|c|}  
 \hline
   $\begin{array}{c} \text{ Hybrid}\\\text{ multiplet}\end{array}$ &
  $\,\,\,\,\,l\,\,\,\,\,$ & $J^{PC}(s=0)$ & $J^{PC}(s=1)$       & $\bm{K}\cdot \hat{\bm{r}}$           \\  \hline\hline
   $H_1$ &
                                  $1$     & $1^{--}$ & $(0,1,2)^{-+}$ & $\Sigma_u^-$, $\Pi_u$ \\
   $H_2$ &
                   $1$     & $1^{++}$ & $(0,1,2)^{+-}$& $\Pi_u$               \\
   $H_3$ &
                   $0$     & $0^{++}$ & $1^{+-}$       & $\Sigma_u^-$          \\
   $H_4$ &
                                   $2$     & $2^{++}$ & $(1,2,3)^{+-}$& $\Sigma_u^-$, $\Pi_u$ \\
   $H_5$ &
                   $2$     & $2^{--}$ & $(1,2,3)^{-+}$ & $\Pi_u$               \\
 \hline
\end{tabular}
\caption{The low-lying hybrid multiplets coming from the $\Sigma_u^-$ and $\Pi_u$ hybrid static energies with $J^{PC}$ quantum numbers ($l\leq2$). The multiplets are ordered by increasing value of the orbital angular momentum.
In Ref.~\cite{Oncala:2017hop}, the multiplets $H_1$, $H_2$, $H_3$, $H_4$ and $H_5$ are named $(s/d)_1$, $p_1$, $p_0$, $(p/f)_2$ and $d_2$, respectively.}
 \label{multiplets}
\end{table}

We split the static energies $E_{\Sigma_u^-,\Pi_u}$ appearing in \eqref{eq:diffeq} into a short-distance part and a long-distance part~\cite{Berwein:2015vca}:
\begin{align}
E_{\Sigma_u^-,\Pi_u}\left(r\right)=
\begin{cases}
V^{\mathrm{RS}}_o(\nu_f)+\Lambda_{\mathrm{RS}}(\nu_f)+b_{\Sigma,\Pi} r^2,&\, r<0.25\,\mathrm{fm}\\
\mathcal{V}_{\Sigma,\Pi}(r), &\, r>0.25\,\mathrm{fm}
\end{cases}\,.
\label{eq:hyb_potential}
\end{align}
For the short-distance part $(r<0.25\, \mathrm{fm})$, we use the RS  octet potential $V^{\mathrm{RS}}_o(r)$ up to order $\alpha^3_s$ in perturbation theory\footnote{
The expression of the RS potential can be found in Appendix B of Ref.~\cite{Berwein:2015vca}.}
and the RS  gluelump mass $\Lambda_{\mathrm{RS}}=0.87$ GeV at the renormalon subtraction scale $\nu_f=1$~GeV \cite{Bali:2003jq,Pineda:2001zq,Pineda:2002se}.
For the long-distance part $\mathcal{V}_{\Sigma,\Pi}(r)$ $(r>0.25\, \mathrm{fm})$, we use~\cite{Berwein:2015vca}
\begin{equation}
\mathcal{V}_{\Sigma,\Pi}(r)=\frac{a_1^{\Sigma,\Pi}}{r}+\sqrt{a_2^{\Sigma,\Pi}\,r^2+a_3^{\Sigma,\Pi}}+a_4^{\Sigma,\Pi}, 
\label{eq:V(r)}
\end{equation}
that smoothly interpolates between the $1/r$ short-distance behaviour and the linear long-distance behaviour.
The parameters $b_{\Sigma,\Pi}$ in Eq.~\eqref{eq:hyb_potential}, and  $a_1^{\Sigma,\Pi}$, $a_2^{\Sigma,\Pi}$, $a_3^{\Sigma,\Pi}$ and $a_4^{\Sigma,\Pi}$ in Eq.~\eqref{eq:V(r)}
depend on the quantum numbers $\Sigma_u^-$ and $\Pi_u$.
They are determined by performing a fit to the lattice data of Refs.~\cite{Juge:2002br, Bali:2003jq}
and demanding that the short-range and the long-range pieces in Eq.~\eqref{eq:hyb_potential} are continuous up to the first derivatives (see Ref.~\cite{Berwein:2015vca} for details).
One obtains 
\begin{align}
a^{\Sigma}_1&=0.000\,\mathrm{GeV \, fm}, & a^{\Sigma}_2&=1.543\,\mathrm{GeV^2/fm^2},& a^{\Sigma}_3 &=0.599\,\mathrm{GeV^2},&a^{\Sigma}_4&=0.154\,\mathrm{GeV}\,,\notag\\
a^\Pi_1&=0.023\,\mathrm{GeV \, fm},&a^\Pi_2&=2.716\,\mathrm{GeV^2/fm^2},&a^\Pi_3&=11.091\,\mathrm{GeV^2},&a^\Pi_4&=-2.536\,\mathrm{GeV}\,,\notag\\
b_{\Sigma}&=1.246\,\mathrm{GeV/fm^2}, &b_{\Pi}&=0.000\,\mathrm{GeV/fm^2}\,.
\label{lattfitpar}
\end{align}
We use for the charm and bottom masses the same RS masses used in Sec.~\ref{subsec:Quarkonium}.
The results for the hybrid spectrum are shown in Table \ref{tab:hybspectrum} and the wavefunctions are shown in appendix~\ref{app:hybridwf}.
The masses for the lowest multiplets have been computed first in~\cite{Berwein:2015vca}; our results agree with and extend those.

\begin{table}[h!]
 \begin{tabular}{|c|c||c||c|}
 \hline
  Multiplet & $J^{PC}$ & $M_{c\bar{c}g}$ & $M_{b\bar{b}g}$ \\
 \hline\hline
 $H_1$  & \multirow{3}{*}{$\{1^{--},(0,1,2)^{-+}\}$} & 4155 & 10786  \\ \cline{3-4}
 $H_1'$  & & 4507  & 10976  \\ \cline{3-4}
 $H_1''$  & & 4812  & 11172  \\ \hline
 $H_2$  & \multirow{3}{*}{$\{1^{++},(0,1,2)^{+-}\}$} & 4286 & 10846  \\ \cline{3-4}
 $H_2'$  & & 4667 & 11060  \\ \cline{3-4}
 $H_2''$  & & 5035 & 11270  \\ \hline
 $H_3$  & \multirow{3}{*}{$\{0^{++},1^{+-}\}$} & 4590 & 11065  \\ \cline{3-4}
 $H_3'$  & & 5054  & 11352  \\ \cline{3-4}
 $H_3''$  & & 5473  & 11616  \\ \hline
 $H_4$ & $\{2^{++},(1,2,3)^{+-}\}$ & 4367 & 10897  \\
 \hline
 $H_5$  & $\{2^{--},(1,2,3)^{-+}\}$ & 4476 & 10948  \\
 \hline
 \end{tabular}
 \caption{Masses of charmonium and bottomonium hybrid states (in MeV) computed using the static potential in Eq.~\eqref{eq:hyb_potential}.
   The hybrid mass is given by $M_{Q\bar{Q}g} = 2m_Q + E_m^{Q\bar{Q}g}$ with $Q=c$, $b$.
   We use the charm, bottom and $1^{+-}$ gluelump masses $m_c^{\mathrm{RS}}= 1.477\,\mathrm{GeV}$, $m_b^{\mathrm{RS}}=4.863\,\mathrm{GeV}$, and $\Lambda_{\mathrm{RS}}=0.87$~GeV, respectively. 
   For the multiplets $H_1$, $H_2$, and $H_3$, the states with a prime and a double prime correspond to first and second excited states. }
\label{tab:hybspectrum}
\end{table}

\section{Hybrid to quarkonium widths}
\label{sec:decay_rate}
Our aim is to compute the semi-inclusive decay rates of a quarkonium hybrid $H_m$ decaying into a quarkonium state $Q_n$: $H_m\rightarrow Q_n + X$, where $X$ denotes light hadrons. 
The energy transfer in the transition $H_m\rightarrow Q_n + X$ is $\Delta E=E_m^{Q\bar{Q}g}-E_{n}^{Q\bar{Q}}$, i.e., the mass difference between the hybrid and the quarkonium. 
In BOEFT, we are integrating out scales up to and including $\Lambda_{\mathrm {QCD}}$, which means that also gluons of energy and momentum $\Delta E$ should be integrated out. 
This leads to an imaginary contribution to the hybrid potential,  
which is related to the semi-inclusive decay width of a hybrid $H_m$ decaying into any quarkonium $Q_n$ by~\cite{Oncala:2017hop}:
\begin{equation}
 \sum_n \Gamma(H_m\to Q_n) = -2\,{\rm Im}\,\langle H_m| V |H_m\rangle\,;
\label{eq:semiinclusivewidth}
\end{equation}
${\rm Im}\,V$ is the imaginary part of the hybrid potential defined in \eqref{eq:LQQg}.
The exclusive decay widths $\Gamma(H_m\to Q_n)$ may be computed by selecting a suitable decay channel in the right-hand side of Eq.~\eqref{eq:semiinclusivewidth}.

We neglect in this study mixing with quarkonium; mixing could however play an important role in the phenomenology 
of quarkonium hybrids whose transition channels are sensibly enhanced or suppressed through the quarkonium component of the physical state~\cite{Oncala:2017hop}.
We restrict to quarkonium states far below the open-flavor threshold.
Furthermore, we restrict to quarkonium and hybrid states for which 
\begin{equation}
\Delta E \gg \Lambda_{\mathrm {QCD}}\,.
\label{eq:DELambda}
\end{equation}
Finally, we require that the emitted gluon cannot resolve the quark-antiquark pair distance, i.e., that the matrix element of the heavy quark-antiquark distance $r$ between the hybrid state and the quarkonium state is smaller than $1/\Delta E$,\footnote{
The matrix element $|\langle Q_n| \bm{r} |H_m\rangle|$ is defined as
$\sqrt{T^{ij}\left(T^{ij}\right)^{\dagger}}$ with $T^{ij}$ given in Eq.~\eqref{eq:matrix_SO}.
}
\begin{equation}
|\langle Q_n|\bm{r}|H_m \rangle| \, \Delta E \ll 1\,.
\label{eq:DEr}
\end{equation}
These conditions, if fulfilled, allow for a treatment of the transition $H_m\to Q_n$
in weakly-coupled pNRQCD, since the gluon at the scale $\Delta E$ is perturbative 
(condition \eqref{eq:DELambda}) and can be multipole expanded (condition \eqref{eq:DEr}).
The explicit computation of the transition $H_m\to Q_n$ in the framework of weakly-coupled pNRQCD is the subject of the remaining of the section.

\subsection{Weakly-coupled pNRQCD}\label{subsec:decay_formulation}
We consider a hybrid decaying into a low-lying quarkonium through the emission of a gluon whose energy satisfies the condition \eqref{eq:DELambda}.
The gluon has enough energy to resolve the heavy quark-antiquark pair in the hybrid and in the quarkonium, and its color configuration.
Therefore the heavy-quark degrees of freedom at the scale $\Delta E$ are quark-antiquark fields, which can be conveniently cast 
into a color singlet field $\S$ and a color-octet field $\Oc$.
They are normalized in color space as $\S=S \mathbb{1}_c/\sqrt{N_c}$ and $\Oc=O^a  T^a/\sqrt{T_F}$,
where $N_c=3$ is the number of colors and $T_F =1/2$ is the normalization of the color matrices.
The quark-antiquark color singlet and octet fields depend in general on the time $t$,
the relative distance, $\bm{r}$, and the center of mass coordinate, $\bm{R}$, of the heavy quark-antiquark pair.
In the short-distance limit, $r\rightarrow 0$, and at leading order in the nonrelativistic expansion, 
the singlet and octet fields are related to the quarkonium field $\Psi$
and the hybrid field ${\Psi}_{\kappa\lambda}$ in Eqs.~\eqref{eq:LQQ} and~\eqref{eq:LQQg} by
\begin{align}
& S\left(\bm{r},\bm{R},t\right) \rightarrow  Z_{\Psi}^{1/2}(\bm{r}) \, \Psi(\bm{r},\bm{R},t) \,, \label{eq:S_Psi}\\
& P^{i\dag}_{\kappa\lambda} O^{a}\left(\bm{r},\bm{R},t\right) G_{\kappa}^{ia}(\bm{R},t) \rightarrow Z_\kappa^{1/2}(\bm{r})\, {\Psi}_{\kappa\la}(\bm{r},\bm{R},t) \,, \label{eq:Oct_hyb}
\end{align}
where
$Z_{\Psi}$ and $Z_{\kappa}$ are normalization factors, 
and $G^{ia}_{\kappa}$ are gluonic fields that match the quantum numbers of the hybrid field on the right-hand side of \eqref{eq:Oct_hyb}.
For low-lying hybrids, the LDF quantum numbers are $\kappa=1^{+-}$;
a gluon field with the same quantum numbers would be the chromomagnetic field $B^{ia} = -\epsilon_{ijk}\,G^{jka}/2$ where $G^{\mu\nu a}$ is the gluon field strength tensor. 
The propagators of weakly interacting quark-antiquark pairs in a color singlet and color octet configuration read in coordinate space 
at leading order in the nonrelativistic and coupling expansion~\cite{Brambilla:1999xf}
\begin{align}
\langle 0 | S\left(\bm{r},\bm{R},t\right)\,S^\dagger\left(\bm{r}',\bm{R}',t'\right)|0\rangle &= \theta(t-t')\, e^{-ih_s(t-t')}\,\delta^3(\bm{r}-\bm{r}')\,\delta^3(\bm{R}-\bm{R}')\,,
\label{eq:SSprop}\\
\langle 0 | O^a\left(\bm{r},\bm{R},t\right)\,O^{b\dagger}\left(\bm{r}',\bm{R}',t'\right)|0\rangle &= \theta(t-t')\, e^{-ih_o(t-t')} \,\phi^{ab}(t,t')\,\delta^3(\bm{r}-\bm{r}')\,\delta^3(\bm{R}-\bm{R}')\,,
\label{eq:OOprop}
\end{align}
where $\phi(t,t')$ is the adjoint static Wilson line, 
\be
\phi(t,t') \equiv {\rm{P}}\exp\left[-ig\int^{t}_{t'}dt\, A_0^{\rm adj}(\bm{R},t)\right]\,,
\label{eq:Wilson}
\ee
P stands for path ordering of the color matrices, and $h_s$ and $h_o$ are the singlet and octet Hamiltonians, respectively, 
\begin{align}
    h_s=-\frac{\bm{\nabla}_r^2}{m_Q} + V_s(r)\,,\qquad 
    h_o=-\frac{\bm{\nabla}_r^2}{m_Q} + V_o(r),
    \label{eq:hs_ho}
\end{align}
$V_s(r) = -C_F \alpha_{\rm s}/r$ and $V_o(r) = \alpha_{\rm s}/(6N_c)$ being the leading-order Coulomb potentials for a color singlet and a color octet state, 
and $C_F = (N_c^2-1)/(2N_c) = 4/3$ the Casimir of the SU(3) fundamental representation.
The potentials $V_s(r)$ and $V_o(r)$ are related to the quarkonium and the hybrid static energies,
$E_{\Sigma_g^+}$, $E_{\Sigma_u^-}$, and $E_{\Pi_u}$, in the short-distance limit, $r\rightarrow 0$, by 
\begin{align}
    E_{\Sigma_g^+}\left(r\right)= V_s(r)+ b_{\Sigma_g^{+}}\,r^2+\dots\,,\qquad
  E_{\Sigma_u^-,\Pi_u}(r) =  V_o(r) + \Lambda + b_{\Sigma,\Pi}r^2+\dots\,,
    \label{eq:static_V_small_r}
\end{align}
where the mass dimension one constant $\Lambda$ is called the gluelump mass and it is related to the correlator of the
suitably normalized gluonic field $G^{ia}_{1^{+-}}$ in the large time $T$ limit by
\begin{equation}
  \langle 0|G^{ia}_{1^{+-}}(\bm{R},T/2)\phi^{ab}(T/2,-T/2)G^{jb}_{1^{+-}}(\bm{R},-T/2)|0\rangle =\delta^{ij} e^{-i\Lambda T}\,.
\label{eq:GGcorrelator}
\end{equation}
The mass dimension three coefficients $b_{\Sigma_g^+}$ and $b_{\Sigma,\Pi}$ are nonperturbative constants to be determined by fitting the lattice data of the static energies
(for the hybrid case see Sec.~\ref{subsec:hybrids}).
Equation~\eqref{eq:static_V_small_r} makes manifest that the static potentials $\Sigma_u^{-}$ and $\Pi_u$ are degenerate at short distances. 

We further assume that the gluon emitted in the $H_m\to Q_n$ transition is not energetic enough to resolve the heavy quark-antiquark distance, see condition \eqref{eq:DEr}.
Under this assumption the gluon field may be multipole expanded in $\bm{r}$ and it is just a function of $t$ and $\bm{R}$.
The leading order chromoelectric-dipole and chromomagnetic-dipole couplings of the gluon with the quark-antiquark pair are encoded
in the Lagrangian $L_{\rm E1}$ and $L_{\rm M1}$, respectively,
\begin{align}  
L_{\rm E1}  &= \int d^3{\bm R} \int d^3{\bm r} \;{\rm Tr}\left(\S^{\dag}\bm{r}\cdot g\bm{E}\,\Oc+\Oc^{\dag}\bm{r}\cdot g\bm{E}\,\S\right)\,,
\label{eq:LpNRQCDE1}\\
L_{\rm M1}  &= \int d^3{\bm R} \int d^3{\bm r} \; \frac{c_F}{m_Q}{\rm Tr}
\left[\S^{\dag}(\bm{S}_1-\bm{S}_2)\cdot g\bm{B}\,\Oc+\Oc^{\dag}(\bm{S}_1-\bm{S}_2)\cdot g\bm{B}\,\S\right]\,.
\label{eq:LpNRQCDM1}
\end{align}  
The trace is over the spin and the color indices, and $c_F$ is a matching coefficient inherited from NRQCD that is 1 at leading order in $\alpha_{\rm s}$.
The field $E^{ia} = G^{i0a}$ is the chromoelectric field.
The matrices ${\bm S}_1= \boldsymbol{\sigma}_1/2$ and ${\bm S}_2= \boldsymbol{\sigma}_2/2$ are the spin of the heavy quark and heavy antiquark, respectively.

\begin{figure}[ht]
\begin{center}
\includegraphics[width=0.3\textwidth,angle=0,clip]{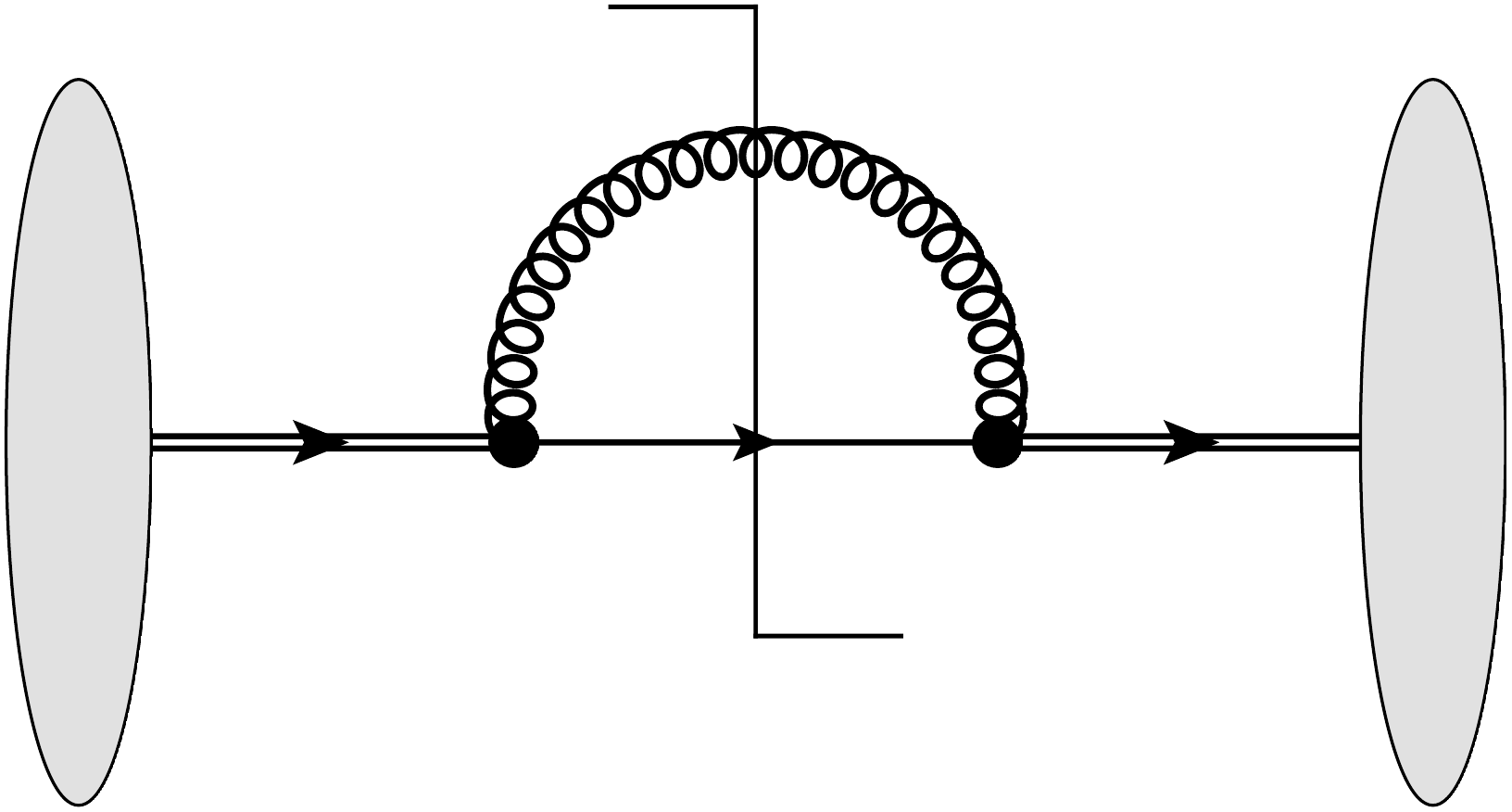}
\caption{One-loop self-energy diagram in pNRQCD.
  The gray blobs represent the hybrid state $H_m$, and the single and double lines represent the $Q\bar{Q}$ pair in the singlet and octet states, respectively.
  The curly line stands for the gluon field and the black dots for pNRQCD vertices.
  The gluonic degrees of freedom that are part of the hybrid are treated as spectator and are not displayed here.
  The vertical line is the cut.}
\label{fig:diagrams}
\end{center}
\end{figure}

The suitable EFT to describe a multipole expanded gluon field interacting with weakly-coupled quark-antiquark pairs, either through chromoelectric or chromomagnetic dipole vertices,
is weakly-coupled pNRQCD~\cite{Pineda:1997bj,Brambilla:1999xf,Brambilla:2004jw,Pineda:2011dg}.
The cut diagram contributing to the $H_m\to Q_n$ transition width at one loop
in weakly-coupled pNRQCD is shown in Fig.~\ref{fig:diagrams}; the gluon carries energy $\Delta E$.

\subsection{Matching and transition rates}
In order to compute the imaginary part of the hybrid potential defined in the BOEFT Lagrangian \eqref{eq:LQQg},
we match the imaginary part of the one-loop two-body Green's function of pNRQCD shown in Fig.~\ref{fig:diagrams} with the corresponding amplitude in BOEFT.
When considering two chromoelectric-dipole vertices from the Lagrangian \eqref{eq:LpNRQCDE1} we obtain an ${\cal O}\left(r^2\right)$ contribution to the potential
responsible for spin-conserving hybrid-to-quarkonium transitions,
whereas when considering two chromomagnetic-dipole vertices from the Lagrangian \eqref{eq:LpNRQCDM1} we obtain an ${\cal O}\left(1/m_Q^2\right)$ contribution
responsible for spin-flipping hybrid-to-quarkonium transitions.
The relative importance of the two processes for hybrid-to-quarkonium transitions, $H_m\to Q_n$,
depends on the relative magnitude of the matrix element $|\langle Q_n|\bm{r}|H_m\rangle|^2$ with respect to $|\langle Q_n|H_m\rangle|^2/m_Q^2$, and 
on the size of the energy gap $\Delta E$ between the hybrid state and the quarkonium state that enters the widths with the third power.

On the weakly-coupled pNRQCD side of the matching, we consider in the large time, $T$, limit the gauge-invariant two-point Green's function in coordinate space
\begin{align}
I_{\lambda\lambda'}(\bm{r},\bm{R},\bm{r}',\bm{R}') \equiv
          \langle 0| P^{i\dag}_{\lambda}G^{ia}(\bm{R},T/2)O^a(\bm{r},\bm{R},T/2) O^{b\dag}(\bm{r}',\bm{R}',-T/2)P^j_{\lambda'}G^{jb}(\bm{R}',-T/2)|0\rangle\,,
\label{eq:I_pNRQCD}
\end{align}
where $P^{i}_{\lambda}$ are the projection operators given in  Eq.~\eqref{eq:P_1},
and the gluonic operators $G^{ia}(\bm{R},t)$ have been introduced in Sec.~\ref{subsec:decay_formulation};
we have dropped again the subscript $\kappa$ as we restrict uniquely to $\kappa=1^{+\,-}$ states.
Repeated color indices $a$, $b$ and spin indices $i$, $j$ are summed.
The two-point Green's function may be expanded in powers of $\bm{r}$ (multipole expansion) or $1/m_Q$ (nonrelativistic expansion):
\begin{equation}
   I_{\lambda\lambda'}(\bm{r},\bm{R},\bm{r}',\bm{R}')=  I^{(0)}_{\lambda\lambda'}(\bm{r},\bm{R},\bm{r}',\bm{R}') +  I^{(2)}_{\lambda\lambda'}(\bm{r},\bm{R},\bm{r}',\bm{R}') + \cdots,
   \label{eq:I_pNRQCD_exp}
\end{equation}
where $I^{(0)}_{\lambda\lambda'}$ is the leading-order (LO) two-point Green's function and $I^{(2)}_{\lambda\lambda'}$ is the next-to-leading-order (NLO) two-point Green's function shown in Fig.~\ref{fig:diagrams}.
The Green's function $I^{(2)}_{\lambda\lambda'}$ develops an imaginary part that is responsible for spin-conserving transitions if the vertices
are chromoelectric-dipole vertices, and for spin-flipping transitions if they are chromomagnetic-dipole vertices.

On the BOEFT side of the matching, the two-point Green's function is given by the large time limit of 
\begin{align}
  &I_{\la\la'}(\bm{r},\bm{R},\bm{r}',\bm{R}') =
    Z^{1/2}(\bm{r})
e^{-i\left( V_{\lambda\lambda'}(r)- P^{i\dag}_{\lambda}\frac{\bnabla^2_r}{m} P^i_{\lambda^{\prime}}\right)T}
    Z^{\dag 1/2}(\bm{r})
  \, \mathbb{1}\,\delta^3(\bm{r}-\bm{r}')\,\delta^3(\bm{R}-\bm{R}')\,,\label{eq:I_hybrid}
\end{align}
where $\mathbb{1}$ is the identity matrix in the spin space of the $Q\bar{Q}$ pair.

\subsubsection{Spin-conserving decay rates}\label{subsec:spin-conserving}
From Eqs.~\eqref{eq:OOprop} and \eqref{eq:GGcorrelator} it follows that the LO two-point function $I^{(0)}_{\lambda\lambda'}$ is given in the large time $T$ limit by 
\begin{equation}
I^{(0)}_{\lambda\lambda'}(\bm{r},\bm{R},\bm{r}',\bm{R}')
=  P^{i\dag}_{\lambda} e^{-i(h_o+\Lambda) T}P^{i}_{\lambda'}\,\mathbb{1}\,\delta^3(\bm{r}-\bm{r}')\,\delta^3(\bm{R}-\bm{R}')\,.\label{eq:I_pNRQCD_0}
\end{equation}
The NLO two-point function $I^{(2)}_{\lambda\lambda'}$ that involves two insertions of the chromoelectric-dipole vertices from the Lagrangian \eqref{eq:LpNRQCDE1} is given in the large time $T$ limit by 
\begin{align}
  &I^{(2)}_{\lambda\lambda'}(\bm{r},\bm{R},\bm{r}',\bm{R}') = -
    {g^2}\frac{T_F}{N_c}
    \int^{T/2}_{-T/2} dt \int^t_{-T/2}dt'\,
 P^{i\dag}_{\lambda}\left[ e^{-ih_o(T/2-t)}r^k e^{-ih_s (t-t')}r^l e^{-ih_o (t'+T/2)}\right] P^{j}_{\lambda'}\nonumber\\
&\qquad\qquad \times\langle 0|G^{ia}(T/2)\phi^{ab}(T/2,t)E^{kb}(t) E^{lc}(t')\phi^{cd}(t',-T/2)G^{jd}(-T/2)|0\rangle\,\mathbb{1}\,\delta^3(\bm{r}-\bm{r}')\,\delta^3(\bm{R}-\bm{R}')\,,
\label{eq:I_pNRQCD_2}
\end{align}
where we have dropped the space coordinates of the fields.

In order to evaluate the two-point function in Eq.~\eqref{eq:I_pNRQCD_2}, we consider the case of energy and momentum flowing into the chromoelectric fields much larger than $\Lambda_{\rm QCD}$, the typical energy and momentum carried by gluon fields $G^{ia}$.
In this situation, we approximate the correlator $\langle 0|G^{ia}(T/2)\phi^{ab}(T/2,t)E^{kb}(t)$ $E^{lc}(t')\phi^{cd}(t',-T/2)G^{jd}(-T/2)|0\rangle$ according to Eq.~\eqref{eq:gluon} of Appendix~\ref{app:matching}.
The evaluation simplifies considerably by taking into account the large time limit.
In the large time limit, we can write
\begin{align}
& \int^{T/2}_{-T/2} dt \int^t_{-T/2}dt'  e^{-ih_o(T/2-t)} (\cdots)  e^{-ih_o (t'+T/2)}  
\nonumber\\
&\hspace{2cm}   =  \left(\int^{0}_{-T/2} dt_1\int^{T+2t_1}_{0}dt_2+\int^{T/2}_{0} dt_1\int^{T-2t_1}_{0}dt_2 \right)  e^{-ih_o(T/2-t_1-t_2/2)} (\cdots)  e^{-ih_o (t_1-t_2/2+T/2)}
 \nonumber\\
&\hspace{2cm} \approx T e^{-ih_oT}  \int^{\infty}_{0} dt_2  \, e^{ih_o t_2/2} (\cdots)  e^{ih_o t_2/2}\,,
\end{align}  
where $t_1\equiv(t+t')/2$, $t_2\equiv t-t'$ and in last line, after using the Baker--Hausdorff lemma, we have retained only the linear term in the large $T$ limit, up to the exponent factor $e^{-ih_oT}$.
The sum of $I^{(0)}_{\lambda\lambda'}$ and $I^{(2)}_{\lambda\lambda'}$ gives
\begin{align}
I_{\lambda\lambda'}(\bm{r},\bm{R},\bm{r}',\bm{R}')
  &=I^{(0)}_{\lambda\lambda'}(\bm{r},\bm{R},\bm{r}',\bm{R}')+I^{(2)}_{\lambda\lambda'}(\bm{r},\bm{R},\bm{r}',\bm{R}') \nonumber\\
  &= P^{i\dag}_{\lambda} e^{-i(h_o+\Lambda) T}
  \left(1  - i \,T \,\Delta V  + \dots \right) P^{i}_{\lambda'} \,\mathbb{1}\,\delta^3(\bm{r}-\bm{r}')\,\delta^3(\bm{R}-\bm{R}')\,,\label{eq:I_pNRQCD_sum}
\end{align}
where the dots stand for terms that are not linear in $T$ and 
\begin{equation}
\Delta V(r)=-\frac{ig^2}{3}\frac{T_F}{N_c}\int^{\infty}_{0}dt\,
e^{ih_o t/2}r^k e^{-ih_s t}r^k e^{ih_o t/2}\int \frac{d^3k}{(2\pi)^3} |\bm{k}|e^{-i|\bm{k}|t}.
\label{eq:DeltaV_matching}
\end{equation}
Considering the definition of $h_o$ given in \eqref{eq:hs_ho}, by equating Eqs.~\eqref{eq:I_pNRQCD_sum} and~\eqref{eq:I_hybrid} we obtain the matching condition 
\begin{equation}
V_{\lambda\lambda'}(r) = P^{i\dag}_{\lambda}V_oP^i_{\lambda^{\prime}} + \Lambda  + P^{i\dag}_{\lambda}\Delta V P^i_{\lambda^{\prime}} \,,
\label{eq:matching_potential}
\end{equation} 
where the sum over the repeated spin index $i$ is implicit.
The form of $V_{\lambda\lambda'}$ agrees with the expression given in Eq.~\eqref{eq:VQQbarg} and Eq.~\eqref{eq:static_V_small_r},
if we identify $\Delta V$ as a contribution of $\mathcal{O}(r^2)$ in the multipole expansion.

Using Eq.~\eqref{eq:Psiwf2}, we can write the spin-conserving decay rate of the hybrid state $|H_m\rangle$ as
\begin{align}
\Gamma=-2\,{\rm Im}\,\langle H_m| \Delta V|H_m\rangle
&= \frac{2g^2}{3}\frac{T_F}{N_c}\, {\rm Re}\int d^3{\bm r}\int^{\infty}_{0}dt\,
\Psi_{(m)}^{i\dagger}(\bm{r}) \left[
e^{ih_o t/2}r^k e^{-ih_s t}r^k e^{ih_o t/2}\int\frac{d^3{\bm k}}{(2\pi)^3} |\bm{k}|e^{-i|\bm{k}|t}\right]\Psi_{(m)}^{i}(\bm{r})\,.
\label{eq:Gamma_spin-conserving}
\end{align}
where  we have defined 
\begin{equation}
    \Psi^{i}_{(m)} \equiv \sum_{\lambda}P^i_{\lambda}\,\Psi^{(m)}_{\lambda},\qquad\qquad \Psi_{\lambda}^{(m)}=P^{i\dag}_{\lambda}\,\Psi^i_{(m)};
  \end{equation}
$i$ is the vector or spin index. 
At this point, if we match the short-distance potentials into the long distance ones, according to the short-distance expansion of Eq.~\eqref{eq:static_V_small_r},
we may promote the singlet and octet Hamiltonians, $h_s$ and $h_o$, to the LO BOEFT quarkonium Hamiltonian $H_\Psi \equiv - \bm{\nabla}^2/m_Q + E_{\Sigma_g^+}(r)$
and hybrid Hamiltonian $H_{\Psi_{1^{+-}}} \equiv - \bm{\nabla}^2/m_Q + E_{\Sigma_u^-,\Pi_u}(r)$ respectively~\cite{Oncala:2017hop,Castella:2021hul},
and the decay rate becomes
\begin{align}
\Gamma
  &= \frac{2g^2}{3}\frac{T_F}{N_c}\,{\rm Re}\int d^3{\bm r}\int^{\infty}_{0}dt\,
  \Psi_{(m)}^{i\dagger}(\bm{r}) \left[e^{iH_{{\Psi}_{1^{+-}}} t/2}r^k e^{-iH_{\Psi} t}r^k e^{i H_{{\Psi}_{1^{+-}}} t/2}
\int\frac{d^3{\bm k}}{(2\pi)^3} |\bm{k}|e^{-i|\bm{k}|t}\right]\Psi_{(m)}^{i}(\bm{r})\,.
\label{eq:Gamma_spin-conserving_SO}
\end{align}
The equation's right-hand side makes manifest that the typical momentum $|\bm{k}|$ in the integral is of the order of the energy gap between the hybrid and the quarkonium,
which is the large energy scale $\Delta E$.
In the case of energy and momentum flowing into the chromoelectric fields of order $\Lambda_{\rm QCD}$,
we would obtain a contribution to the hybrid potential still of ${\cal O}\left(r^2\right)$ in the multipole expansion but suppressed by $(\Lambda_{\rm QCD}/\Delta E)^3$ relative to $\Delta V$.
After using the completeness relation for the quarkonium eigenfunctions 
$\Phi^{Q\bar{Q}}_{(n)}$, 
we obtain the semi-inclusive decay rate of the process $H_m\rightarrow Q_n + X$ for each intermediate quarkonium state $Q_n$,
\begin{align}
\Gamma(H_m\to Q_n)=\frac{4\,\alpha_{\rm s}\left(\Delta E\right)\, T_F}{3N_c}\,T^{ij}\,(T^{ij})^\dag\,\Delta E^3\,,
\label{eq:Gamma_spincons}
\end{align}
where $\Delta E= E_m^{Q\bar{Q}g}-E^{Q\bar{Q}}_n$ is the energy difference and 
\begin{align}
  T^{ij}&\equiv
          \int d^3{\bm r}\,\Psi_{(m)}^{i\dagger}(\bm{r})\,r^{j}\,\Phi^{Q\bar{Q}}_{(n)}(\bm{r})\,.
\label{eq:matrix_SO}
\end{align}
We have also made explicit that the natural scale of $\alpha_{\rm s}$ is $\Delta E$.

The decay rate in Eq.~\eqref{eq:Gamma_spincons} has been also derived in Refs.~\cite{Oncala:2017hop} and~\cite{Castella:2021hul}. 
However, in Ref.~\cite{Oncala:2017hop} only the diagonal matrix elements $T^{ii}$
were included in the decay rate in Eq.~\eqref{eq:Gamma_spincons}.
If we decompose $T^{ij}$ as 
\begin{align}
T^{ij}=T_0\,\delta^{ij}+T_1^{ij}+T_2^{ij}\,,
\end{align}
with
\begin{align}
T_0&\equiv \frac{1}{3}\,T^{ll}\,,\qquad
T_1^{ij}\equiv \frac{1}{2}(T^{ij}-T^{ji}),\qquad
T_2^{ij}\equiv \frac{1}{2}(T^{ij}+T^{ji})-\frac{\delta^{ij}}{3}T^{ll},
\end{align}
then, we see that
\begin{align}
T^{ij}(T^{ij})^{\dagger}& =\frac{1}{3}T^{ii}(T^{ii})^{\dagger}+T_1^{ij}(T_1^{ij})^{\dagger}+T_2^{ij}(T_2^{ij})^{\dagger}\,.
\end{align}
The result in Ref.~\cite{Oncala:2017hop} is equivalent to setting $T^{ij}_1=T^{ij}_2=0$ and multiplying the term $T^{ii}(T^{ii})^{\dagger}/3$ by 3;
this leads to a selection rule that hybrids with $L=L_{Q\bar{Q}}$ do not decay.
We will see that by accounting for the full tensor structure of the matrix element $T^{ij}$ in Eq.~\eqref{eq:Gamma_spincons},
also decays of hybrids with  $L=L_{Q\bar{Q}}$ turn out to be possible.

In spin-conserving decays, the spin of the $Q\bar{Q}$ pair in the hybrid and in the final state are the same:
the non-vanishing of the matrix element~\eqref{eq:matrix_SO} constrains spin-$0$ hybrids to decay into spin-$0$ final states and spin-$1$ hybrids to decay into spin-$1$ final states.
For the spin-$1$ hybrid states, the spin-conserving rate in Eq.~\eqref{eq:Gamma_spincons} is multiplied by a factor $3$ corresponding to the $3$ polarizations of the spin-triplet final quarkonium state.

\subsubsection{Spin-flipping decay rates}\label{subsec:spin-flipping}
The chromomagnetic-dipole interaction 
in the Lagrangian~\eqref{eq:LpNRQCDM1} is responsible for spin-flipping decays of hybrid to quarkonium (spin-$0$ hybrid decaying to spin-$1$ quarkonium and vice versa).
Spin-flipping transition widths are, in principle, suppressed by powers of the heavy-quark mass due to the heavy-quark spin symmetry;
however, as we already remarked, if they turn out to be actually smaller than spin-conserving transition widths depends on the relative size of $|\langle Q_n|H_m\rangle|/m_Q$ with respect to $|\langle Q_n|\bm{r}|H_m\rangle|$, which are not related by power counting in an obvious manner, 
and on the size of the energy gap between hybrid state and quarkonium state.
The matching of the imaginary part of the hybrid BOEFT potential goes exactly in the same way as in the previous section, 
with the chromoelectric-dipole term $\bm{r}\cdot\bm{E}$ replaced by the chromomagnetic one $(\bm{S}_1-\bm{S}_2)\cdot\bm{B}/m_Q$.\footnote{
At tree level it holds that $\langle 0|E^{ia}(t) E^{ib}(t')|0\rangle = \langle 0|B^{ia}(t) B^{ib}(t')|0\rangle$.}
The spin-flipping transition rate is given by Eq.~\eqref{eq:Gamma_spincons} with $T^{ij}$ now 
\begin{align}
  T^{ij}&\equiv
       \frac{1}{m_Q}   \left[\int d^3{\bm r}\,\Psi_{(m)}^{i\dagger}(\bm{r})\,
\Phi^Q_{(n)}(\bm{r})\right]\langle \chi_H|\left(S_1^{ j}-S_2^{j}\right)|\chi_Q\rangle\,,
\label{eq:matrix_SO-spin}
\end{align}
where  ${\bm S}_1$ and ${\bm S}_2$ are the spin vectors of the heavy quark and heavy antiquark and $|\chi_H\rangle$ and $|\chi_Q\rangle$ denote the hybrid and quarkonium spin states, respectively.
The spin-matrix elements are computed in Appendix~\ref{app:spin-matrix}. The expression for the spin-flipping transition rate agrees with the one found in Ref.~\cite{Castella:2021hul}.

In the spin-flipping decays, the spin of the $Q\bar{Q}$ pair in the hybrid and in the final state are different:
the non-vanishing of the matrix element~\eqref{eq:matrix_SO-spin} constrains spin-$0$ hybrids to decay into spin-$1$ final states and spin-$1$ hybrids to decay into spin-$0$ final states.
For the spin-$0$ hybrid states, the spin-flipping rate in Eq.~\eqref{eq:Gamma_spincons} is multiplied by a factor $3$
corresponding to the $3$ polarizations of the spin-triplet final quarkonium state.

\section{Results and comparison with experiments}\label{sec:results}

\subsection{Exotic XYZ states and hybrids}\label{subsec:exp}
\begin{figure}[ht]
\begin{center}
\includegraphics[width=0.65\textwidth,angle=0,clip]{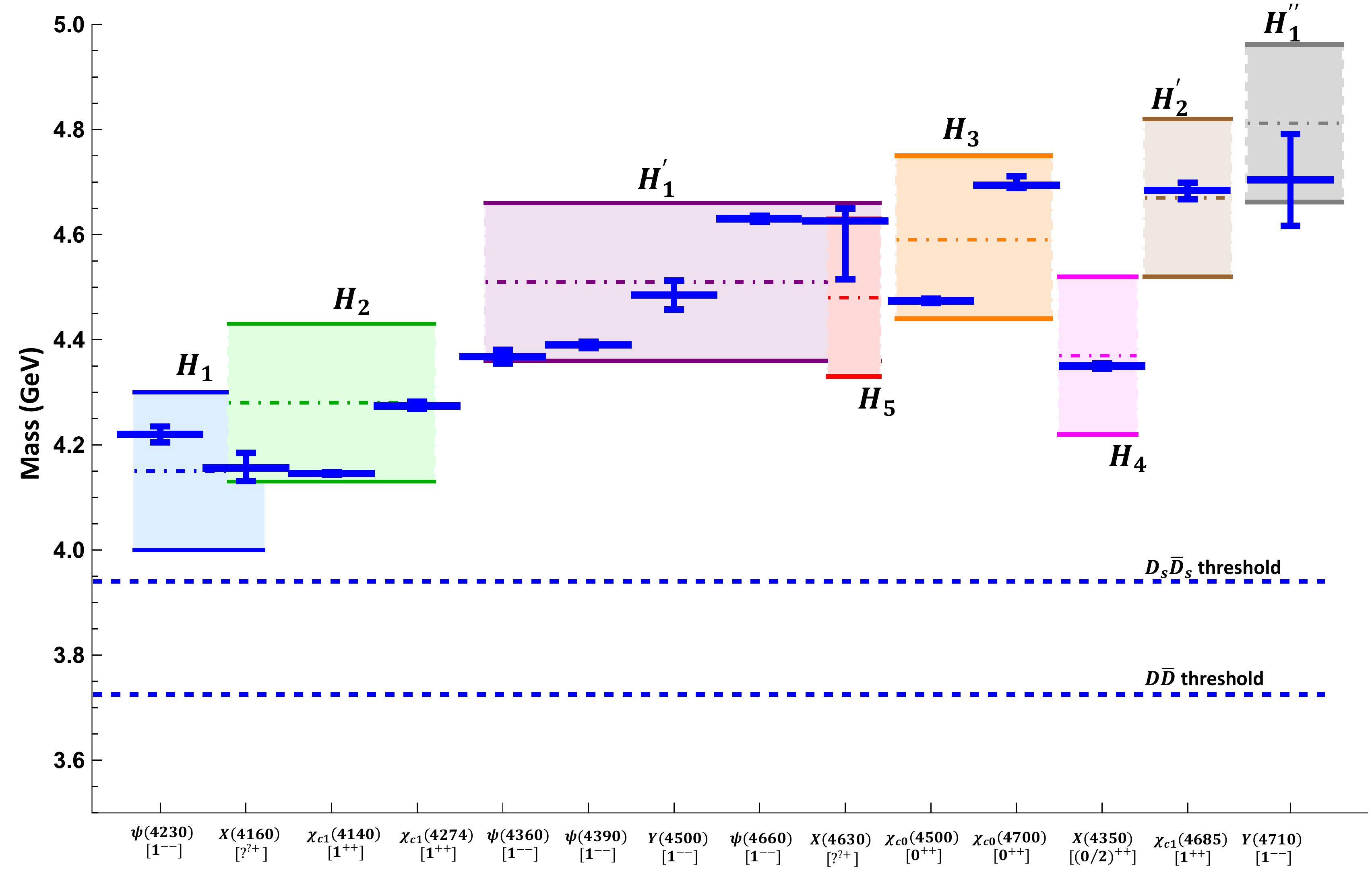}
\caption{Comparison of the mass spectrum of the neutral exotic charmonium-like  states shown in Table~\ref{tab:exotic}
  with results for hybrids obtained by solving the coupled  Schr\"{o}dinger equations~\eqref{eq:diffeq}.
  The experimental states are represented by horizontal solid blue lines with vertical error bars.
  Our results for the multiplets $H_1$, $H_1'$, $H_1''$, $H_2$, $H_2'$, $H_3$, $H_4$ and $H_5$ are plotted with error bands corresponding to a gluelump mass uncertainty of $\pm0.15$~GeV.
  The figure has been adapted and updated from Ref.~\cite{Brambilla:2019esw}.}
\label{fig:cchybrids}
\end{center}
\end{figure}

\begin{figure}[h]
\begin{center}
\includegraphics[width=0.65\textwidth,angle=0,clip]{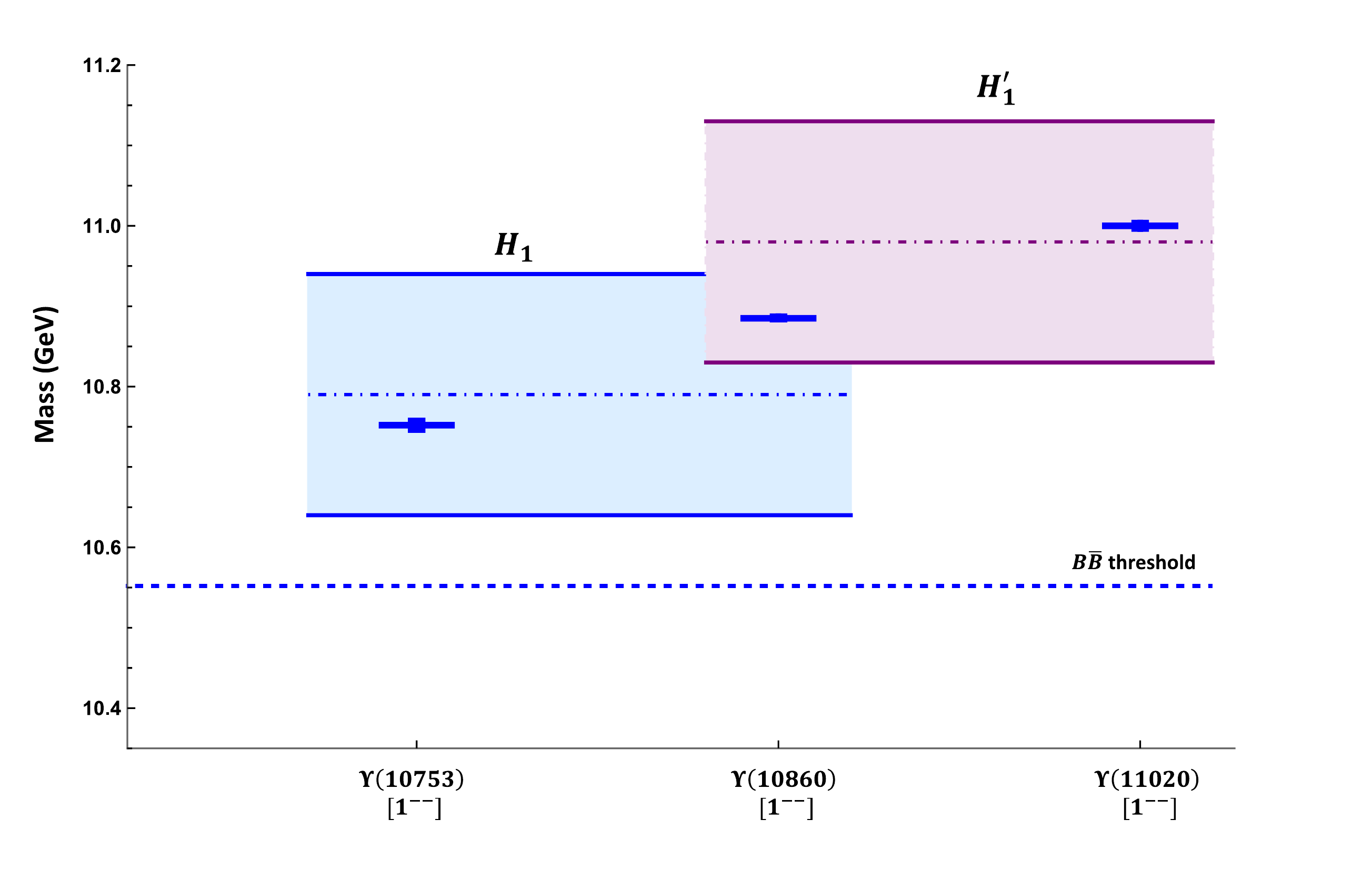}
\caption{Comparison of the mass spectrum of the neutral exotic bottomonium-like states shown in Table~\ref{tab:exotic}
  with results for hybrids obtained by solving the coupled  Schr\"{o}dinger equations~\eqref{eq:diffeq}.
  The experimental states are represented by solid blue lines with vertical error bars.
  Our results for the multiplets $H_1$ and  $H_1'$  are plotted with error bands corresponding to the gluelump mass uncertainty of $\pm0.15$~GeV.
  We only show the multiplets $H_1$ and $H_1'$ as there are only three exotic states with matching quantum numbers (see Table~\ref{tab:exotic}).}
\label{fig:bbhybrids}
\end{center}
\end{figure}

The heavy-quark hybrid states are isoscalar neutral mesons.
The list of XYZ exotic states that are potential candidates for heavy-quark hybrids are the neutral heavy-quark mesons above the open flavor threshold.
An updated list  of such  states can be found in Table~\ref{tab:exotic}~\cite{Workman:2022ynf}.
Several of the exotic states in Table~\ref{tab:exotic} have quantum numbers $1^{--}$ and $0^{++}$ or $2^{++}$ as they are generally observed
in the production channels of $e^{+}e^{-}$ or $\gamma\gamma$ annihilation.
After matching the quantum numbers $J^{PC}$ of the hybrids in Table~\ref{tab:hybspectrum} with the XYZ states in Table~\ref{tab:exotic},
potential XYZ  candidates for charmonium and bottomonium hybrids are shown in Figs.~\ref{fig:cchybrids} and \ref{fig:bbhybrids}, respectively.
The bands in Figs.~\ref{fig:cchybrids} and \ref{fig:bbhybrids} represent only the uncertainty in the mass of the hybrids
due to the uncertainty in the gluelump mass, $\Lambda_{\mathrm{RS}} = 0.87\pm0.15$~GeV.  

\begin{table}[h!]
\begin{center}
\small{\renewcommand{\arraystretch}{0.98}
\scriptsize
\begin{tabular}{|cccccc|}
\hline
   $\begin{array}{c} {\rm State}\\({\rm PDG})\end{array}$ & $\begin{array}{c} {\rm State}\\({\rm Former})\end{array}$      &   $M$~(MeV)       & $\Gamma$~(MeV) &  $J^{PC}$    &  Decay modes                            \\
   \hline
 $\chi_{c1}\left(4140\right)$ & $X(4140)$ & $4146.5\pm 3.0$ & $19^{+7}_{-5}$ & $1^{++}$ & $\phi\,J/\psi$\\
  
 $X\left(4160\right)$ &  & $4153^{+23}_{-21}$ & $136^{+60}_{-35}$ & $?^{??}$ & 
 $\phi\,J/\psi $, $D^*\bar{D}^*$\\
 
$\psi\left(4230\right)$ & $Y(4230)$ & $4222.7\pm 2.6$ & $49\pm 8$ & $1^{--}$ & $\pi^+\pi^-\,J/\psi$, $\omega\,\chi_{c0}(1P),$\\
& $Y(4260)$ & & & & $\pi^+\pi^-h_c(1P)$\\
  
$\chi_{c1}\left(4274\right)$ & $Y(4274)$ & $4286^{+8}_{-9}$ & $51\pm 7$ & $1^{++}$ & $\phi\,J/\psi$\\
 
$X\left(4350\right)$ &  & $4350.6^{+4.7}_{-5.1}$ & $13^{+18}_{-10}$ & $\left(0/2\right)^{++}$ &  $\phi\,J/\psi$\\

$\psi\left(4360\right)$ & $Y(4360)$ & $4372\pm 9$ & $115\pm13$ & $1^{--}$ & $\pi^+\pi^-J/\psi$,\\
&$Y(4320)$&&&&
$\pi^+\pi^-\psi(2S)$\\

$\psi\left(4390\right)$\footnote{This state is not listed in~\cite{Workman:2022ynf}. 
Its existence has been suggested in the BESIII analysis of Ref.~\cite{BESIII:2016adj}. For a critical review see Ref.~\cite{Brambilla:2019esw}.}  & $Y(4390)$ & $4390\pm 6$ & $139^{+16}_{-20}$ & $1^{--}$ & $\eta J/\psi$, $\pi^+\pi^-h_c(1P)$\\

$\chi_{c0}\left(4500\right)$ & $X(4500)$ & $4474\pm4$ & $77^{+12}_{-10}$ & $0^{++}$ & $\phi\,J/\psi$\\

$Y\left(4500\right)$\footnote{State recently observed by the BESIII collaboration~\cite{BESIII:2022joj}} &  & $4484.7\pm 27.5$ & $111 \pm 34$ & $1^{--}$ &   \\

$X\left(4630\right)$ \footnote{State recently observed by the LHCb collaboration~\cite{LHCb:2021uow}} &  & $4626^{+24}_{-111}$ & $174^{+137}_{-78}$ & $?^{?+}$ &  $\phi\,J/\psi$\\

$\psi\left(4660\right)$ & $Y(4660)$ & $4630\pm 6$ & $72^{+14}_{-12}$ & $1^{--}$ & $\pi^+\pi^-\psi(2S)$, $\Lambda_c^{+}\bar{\Lambda}_c^{-},$\\
& $X(4660)$ & & & & $D_s^{+}D_{s1}(2536)$\\

$\chi_{c1}\left(4685\right)$ \footnote{State recently observed by the LHCb collaboration~\cite{LHCb:2021uow}} &  & $4684^{+15}_{-17}$ & $126^{+40}_{-44}$ & $1^{++}$ & $\phi\,J/\psi$\\

$\chi_{c0}\left(4700\right)$ & $X(4700)$ & $4694^{+17}_{-5}$ & $87^{+18}_{-10}$ & $0^{++}$ & $\phi\,J/\psi$\\

$Y\left(4710\right)$\footnote{State recently observed by the BESIII collaboration~\cite{Ablikim:2022yav}} &  & $4704\pm 87$ & $183 \pm 146$ & $1^{--}$ &   \\
\hline
$\Upsilon\left(10753\right)$ &  & $10752.7^{+5.9}_{-6.0}$ & $36^{+18}_{-12}$ & $1^{--}$ & $\pi\pi\Upsilon\left(1S,2S,3S\right)$\\

$\Upsilon\left(10860\right)$ & $\Upsilon\left(5S\right)$ & $10885.2^{+2.6}_{-1.6}$ & $37\pm 4$ & $1^{--}$ & $\pi\pi\Upsilon\left(1S,2S,3S\right)$,\\
&&&&&$\pi^+\pi^{-}h_b\left(1P, 2P\right)$,\\
&&&&&$\eta\Upsilon(1S, 2S)$, $\pi^+\pi^{-}\Upsilon\left(1D\right)$\\
&&&&& (see PDG listings)\\

$\Upsilon\left(11020\right)$ & $\Upsilon\left(6S\right)$ & $11000\pm 4$ & $24^{+8}_{-6}$ & $1^{--}$ & $\pi\pi\Upsilon\left(1S,2S,3S\right)$,\\
&&&&&$\pi^+\pi^{-}h_b\left(1P, 2P\right)$,\\
&&&&& (see PDG listings)\\
 \hline
\end{tabular}
\caption{The isoscalar neutral meson states ordered by mass above the open flavor thresholds in the $c\bar{c}$ and $b\bar{b}$ regions.
  Following Refs.~\cite{Workman:2022ynf, Olsen:2017bmm, Yuan:2021wpg, Brambilla:2019esw}, we have only included states that are possible candidates for hybrid states.
  The second column reports the old names still used in the literature. The table has been adapted from Ref.~\cite{Workman:2022ynf}. }
\label{tab:exotic}}
\end{center}
\end{table}

In the charmonium sector, the first exotic $\psi$ state, the $\psi(4260)$ (also know as $Y(4260)$),
was observed by the BaBar experiment in the process $e^{+}e^{-}\rightarrow \pi^+\pi^{-} J/\psi$ \cite{BaBar:2005hhc}.
Later, precise measurements of the $e^{+}e^{-}\rightarrow \pi^+\pi^{-} J/\psi$ cross sections by the BESIII experiment reported that the $\psi(4260)$ state actually has a lower mass that is more consistent with the state $\psi(4230)$ \cite{BESIII:2016bnd}.
Additionally, the BESIII experiment also reported a new resonance with a mass of around $4.32\,\mathrm{GeV}$
that is observed as a distinct shoulder on the high-mass side of the $\psi(4260)$ peak.
This new resonance was named $\psi(4320)$ (also know as $Y(4320)$).
Since, both mass and width of the $\psi(4320)$ are consistent with those of the $\psi(4360)$  resonance observed in $e^{+} e^{-}\rightarrow \pi^+\pi^{-} \psi(2S)$
by BaBar and Belle \cite{BaBar:2012hpr,Belle:2014wyt}, 
they could be the same state.
So, there are only four confirmed states\footnote{
  The exotic state $Y(4008)$ has not been confirmed by other experiments such as BESIII and BaBar \cite{BESIII:2016bnd, BaBar:2012vyb}. }
with quantum numbers $J^{PC}=1^{--}$: $\psi(4230)$, $\psi(4360)$, $\psi(4390)$, and $\psi(4660)$ \cite{Workman:2022ynf, Olsen:2017bmm, Yuan:2021wpg, Brambilla:2019esw}.
The quantum numbers $J^{PC}=1^{--}$ correspond to the spin-singlet members $1^{--}$ of the hybrid $H_1$ multiplet.
The $\psi(4230)$ state falls in the range of masses for the charmonium hybrids belonging to the $H_1$ multiplet, while the states $\psi(4360)$, $\psi(4390)$,
and $\psi(4660)$ have a mass that is compatible with the excited spin singlet states belonging to the $H_1^{'}$ multiplet after including the uncertainties in the gluelump mass.
From Table~\ref{tab:exotic}, we see that the states $\psi(4230)$ and $\psi(4390)$ decay both to the spin singlet charmonium, $h_c(1P)$, and to the spin triplet charmonium, $J/\psi$. This could be consistent with hybrid spin-conserving and spin-flipping decays, respectively.
Instead, the states $\psi(4360)$ and $\psi(4660)$ have only been observed to decay to spin triplet charmonium states, $J/\psi$ and $\psi(2S)$.
Recently, the BESIII collaboration has suggested the existence of two possible new states with quantum numbers $J^{PC}=1^{--}$, $Y(4500)$ and $Y(4710)$, from resonance structures in the $e^{+} e^{-}\rightarrow K^+K^{-} J/\psi$ and $e^{+} e^{-}\rightarrow K_S^0K^0_S J/\psi$ cross sections, respectively~\cite{BESIII:2022joj, Ablikim:2022yav}.
The masses and the quantum numbers of these states are compatible with the excited spin singlet $H_1^{'}$ and $H_1^{''}$ hybrid multiplets after including the uncertainties from the gluelump mass.

The quantum numbers $J^{PC}=1^{++}$  and the mass of the $\chi_{c1}(4140)$ and $\chi_{c1}(4274)$ suggest that they could be candidates for
the spin singlet $1^{++}$ member of the $H_2$ hybrid multiplet within uncertainties.
For the spin singlet member of the $H_2$ multiplet, a spin-conserving decay leads to a spin singlet $\eta_c(1S)$ quarkonium in the final state and a spin-flipping decay leads to a spin triplet $\chi_c(1P)$ quarkonium in the final state. 
The states $\chi_{c1}(4140)$ and $\chi_{c1}(4274)$, however, have been observed to decay 
only to $\phi\,J/\psi$. 
It has been suggested that these states could be  isospin-$0$ charmonium tetraquark states~\cite{Braaten:2014qka, Giron:2020qpb}.
The $J^{PC}$ quantum numbers of the $X(4160)$ have not yet been determined. 
A positive charge conjugation and the mass could make  
it a candidate for the spin triplet $\left(0, 1, 2\right)^{-+}$ member of the $H_1$ multiplet or the spin singlet $1^{++}$ member of the $H_2$ multiplet.
Recently, the LHCb collaboration reported two new exotic states, $X(4630)$ and $\chi_{c1}\left(4685\right)$, with quantum numbers $J^{PC}=?^{?+}$ and $J^{PC}=1^{++}$
in the $B^{+} \rightarrow J/\psi \phi K^+ $ decay~\cite{LHCb:2021uow}. 
The favoured quantum numbers for $X(4630)$ are $J^{PC}=\left(1 \, \mbox{or} \, 2\right)^{-+}$ \cite{Workman:2022ynf, LHCb:2021uow}. Based on the   
quantum numbers and mass, the $X(4630)$ state  could be a candidate for  the excited spin triplet $\left(0, 1, 2\right)^{-+}$ member of the $H_1$ multiplet or the spin triplet $\left(1, 2, 3\right)^{-+}$ member of the $H_5$ multiplet after including the uncertainties from the gluelump mass.
The quantum numbers $1^{++}$ and the mass of $\chi_{c1}\left(4685\right)$ are compatible with the spin singlet state of the excited $H_2^{'}$ multiplet after accounting for the uncertainties from the gluelump mass.
For the $X(4630)$ and $\chi_{c1}\left(4685\right)$, only the decay to $\phi\,J/\psi$ has been seen until now. 

The quantum numbers of $X(4350)$ are $J^{PC}=(0 \, \mbox{or} \, 2)^{++}$~\cite{Belle:2009rkh}.
The mass of the $X(4350)$ suggests that it could be a candidate for the spin singlet $2^{++}$ member of the $H_4$ multiplet.
The quantum numbers $J^{PC}= 0^{++}$ and the masses of the $\chi_{c0}(4500)$ and $\chi_{c0}(4700)$ suggest that they could be candidates for
the spin singlet $0^{++}$ member of the $H_3$ hybrid multiplet within uncertainties.
For the spin singlet member of the $H_3$ and $H_4$ multiplets,
the spin-conserving transitions lead to the spin singlet $\eta_c(1S)$ quarkonium in the final state and the spin-flipping transitions lead to the spin triplet $\chi_c(1P)$ quarkonium in the final state.
However, the states $X(4350)$, $\chi_{c0}(4500)$, and $\chi_{c0}(4700)$ 
have been observed to decay only to $\phi\,J/\psi$.

In the bottomonium sector, there are only three exotic candidates for the hybrid states with  quantum numbers $J^{PC}=1^{--}$: $\Upsilon(10753)$, $\Upsilon(10860)$, and $\Upsilon(11020)$.
The quantum number $1^{--}$ corresponds to the spin singlet member $1^{--}$ of the bottomonium  hybrid $H_1$ multiplet or its excitation.
The mass of the $\Upsilon(10753)$ and $\Upsilon(11020)$ states suggests that they could be identified with states in the $H_1$ or $H_1^{'}$ multiplets, respectively.
The mass of the $\Upsilon(10860)$, besides being consistent with a conventional $\Upsilon(5S)$ bottomonium state,
is compatible with both $H_1$ and $H_1^{'}$ bottomonium hybrid multiplets within uncertainties.
From Table~\ref{tab:exotic}, we notice that the states $\Upsilon(10860)$ and $\Upsilon(11020)$ decay both to the spin singlet bottomonium state $h_b(1P)$
and to the spin triplet bottomonium states $\Upsilon(nS)$.
The decay to $h_b(1P)$ could correspond to a spin-conserving transition and the decay to $\Upsilon(nS)$ could correspond to a spin-flipping transition.
The state  $\Upsilon(10753)$ has been observed to decay only to spin triplet $\Upsilon(nS)$ bottomonium states.
Recent studies have suggested that some of these states could be conventional quarkonium or tetraquark
states~\cite{Bruschini:2018lse, Bicudo:2020qhp, Liang:2019geg, Giron:2020qpb, Li:2019qsg, Wang:2019veq, Chen:2019uzm, Ali:2019okl}.

\begin{table}[htb]
\begin{center}
\small{\renewcommand{\arraystretch}{1.2}
\scriptsize
\begin{tabular}{|c||c|c||c||c|}
\hline
$H_m\left[J^{PC}\,\right]\,\left(\mathrm{Mass}\right)\longrightarrow Q_n \left[J^{PC}\,\right]$  & $\Gamma$~(MeV)\\
\hline
\multicolumn{2}{|c|} {\hspace{0.3 cm} Charmonium hybrid}\\
\hline
$H_2\left[\,1^{++}\,\right]\,\left(4667\right) \longrightarrow \eta_c\left(1S\right)\left[\,0^{-+}\,\right]$    & 65 $^{+27}_{-14}$ $^{+20}_{-17}$ \\

$H_2\left[\,1^{++}\,\right]\,\left(5035\right) \longrightarrow \eta_c\left(1S\right)\left[\,0^{-+}\,\right]$    &  31 $^{+11}_{-6}$ $^{+8}_{-7}$ \\

$H_2\left[\,1^{++}\,\right]\,\left(5035\right) \longrightarrow \eta_c\left(2S\right)\left[\,0^{-+}\,\right]$    &  45 $^{+20}_{-10}$ $^{+16}_{-13}$ \\
\hline

$H_3\left[\,0^{++}\,\right]\,\left(5054\right) \longrightarrow \eta_c\left(1S\right)\left[\,0^{-+}\,\right]$  &  45 $^{+16}_{-9}$ $^{+11}_{-9}$  \\

$H_3\left[\,0^{++}\,\right]\,\left(5473\right) \longrightarrow \eta_c\left(1S\right)\left[\,0^{-+}\,\right]$   & 18 $^{+6}_{-3}$ $^{+4}_{-3}$  \\

$H_3\left[\,0^{++}\,\right]\,\left(5473\right) \longrightarrow \eta_c\left(2S\right)\left[\,0^{-+}\,\right]$  & 26 $^{+10}_{-5}$ $^{+7}_{-6}$   \\
\hline
\multicolumn{2}{|c|} {\hspace{0.3 cm} Bottomonium hybrid}\\
\hline

$H_1\left[\,1^{--}\,\right]\,\left(10976\right) \longrightarrow h_b\left(1P\right)\left[\,1^{+-}\,\right]$   & 15 $^{+8}_{-4}$ $^{+7}_{-5}$  \\

$H_1\left[\,1^{--}\,\right]\,\left(11172\right) \longrightarrow h_b\left(2P\right)\left[\,1^{+-}\,\right]$   & 22 $^{+14}_{-6}$ $^{+13}_{-9}$ \\

\hline
$H_2\left[\,1^{++}\,\right]\,\left(10846\right) \longrightarrow \eta_b\left(1S\right)\left[\,0^{-+}\,\right]$   & 29 $^{+13}_{-7}$ $^{+10}_{-8}$ \\

$H_2\left[\,1^{++}\,\right]\,\left(11060\right) \longrightarrow \eta_b\left(1S\right)\left[\,0^{-+}\,\right]$    & 28 $^{+11}_{-6}$ $^{+9}_{-7}$ \\

$H_2\left[\,1^{++}\,\right]\,\left(11060\right) \longrightarrow \eta_b\left(2S\right)\left[\,0^{-+}\,\right]$   & 0.22 $^{+0.12}_{-0.06}$ $^{+0.11}_{-0.08}$ \\

$H_2\left[\,1^{++}\,\right]\,\left(11270\right) \longrightarrow \eta_b\left(1S\right)\left[\,0^{-+}\,\right]$    & 22 $^{+8}_{-4}$ $^{+6}_{-5}$ \\

$H_2\left[\,1^{++}\,\right]\,\left(11270\right) \longrightarrow \eta_b\left(2S\right)\left[\,0^{-+}\,\right]$   & 6 $^{+3}_{-1}$ $^{+2}_{-2}$ \\

$H_2\left[\,1^{++}\,\right]\,\left(11270\right) \longrightarrow \eta_b\left(3S\right)\left[\,0^{-+}\,\right]$    & 3 $^{+2}_{-1}$ $^{+2}_{-1}$ \\
\hline
$H_3\left[\,0^{++}\,\right]\,\left(11065\right) \longrightarrow \eta_b\left(1S\right)\left[\,0^{-+}\,\right]$  &  69 $^{+28}_{-15}$ $^{+21}_{-17}$  \\

$H_3\left[\,0^{++}\,\right]\,\left(11352\right) \longrightarrow \eta_b\left(1S\right)\left[\,0^{-+}\,\right]$  &   34 $^{+12}_{-7}$  $^{+9}_{-7}$ \\

$H_3\left[\,0^{++}\,\right]\,\left(11352\right) \longrightarrow \eta_b\left(2S\right)\left[\,0^{-+}\,\right]$  &  42 $^{+19}_{-10}$  $^{+16}_{-13}$  \\

$H_3\left[\,0^{++}\,\right]\,\left(11616\right) \longrightarrow \eta_b\left(1S\right)\left[\,0^{-+}\,\right]$   & 19 $^{+6}_{-4}$  $^{+4}_{-4}$ \\

$H_3\left[\,0^{++}\,\right]\,\left(11616\right) \longrightarrow \eta_b\left(2S\right)\left[\,0^{-+}\,\right]$   & 20 $^{+8}_{-4}$  $^{+6}_{-5}$\\
\hline
\end{tabular}
\caption{Spin-conserving semi-inclusive decay rates of hybrids decaying to quarkonia below threshold, 
due to the chromoelectric-dipole interaction~\eqref{eq:LpNRQCDE1}.
The decay rates are computed from Eqs.~\eqref{eq:Gamma_spincons} and \eqref{eq:matrix_SO}.
The hybrid states are denoted by $H_m\left[J^{PC}\,\right]\,\left(\mathrm{mass}\right)$, 
where $J$ is the total angular momentum quantum number including the spin, and the masses are in MeV.
The quarkonium states are denoted by the physical states. 
For the quarkonium states, we use the spin-averaged masses given in Table~\ref{tab:QQbarspectrum}. The first error comes from varying the scale of $\alpha_{\rm s}$ from $\Delta E/2$ to $2\,\Delta E$,
and the second one from the gluelump mass uncertainty of $\pm\, 0.15\,\mathrm{GeV}$.
We only show decay rates for which  $\Delta E \gtrsim 0.8$~GeV , $\alpha_{\rm s}(\Delta E) \lesssim 0.4$ and $|\langle Q_n| \bm{r} |H_m\rangle| \, \Delta E \lesssim 0.8$. For this last condition see Table~\ref{tab:rDeltaE}.}
\label{tab:Gamma_Exclusive_spin-conserving}
}
\end{center}
\end{table}

\subsection{Results for the decay rates }\label{subsec:decay_results}
The exotic XYZ states in the charmonium sector shown in Fig.~\ref{fig:cchybrids} have mostly quantum numbers $J^{PC}=$ $1^{--}$, $1^{++}$,  $0^{++}$  and $2^{++}$
that correspond to the $J^{PC}$
quantum numbers of the spin-singlet members of the hybrid multiplets $H_1\left[1^{--}\right]$, 
$H_2\left[1^{++}\right]$, $H_3\left[0^{++}\right]$, $H_4\left[2^{++}\right]$ and their excitations.
The exotic state $X(4630)$ could have $J^{PC}$ quantum numbers $1^{-+}$ or $2^{-+}$, and be a spin triplet member of the hybrid multiplets $H_1\left[(0,1,2)^{-+} \right]$
or $H_5\left[(1,2,3)^{-+} \right]$.
The exotic XYZ states in the bottomonium sector shown in Fig.~\ref{fig:bbhybrids} have quantum numbers $J^{PC}=$ $1^{--}$
that correspond to the $J^{PC}$ 
quantum numbers of the spin-singlet members of the hybrid multiplet $H_1\left[1^{--}\right]$ and its excitations. In the following, we focus solely on these hybrid states and compute the semi-inclusive spin-conserving and spin-flipping transition rates to quarkonia. 
The spin-conserving decays of hybrids to quarkonia, $H_m \rightarrow Q_n+X$, where $X$ denotes light hadrons, are induced by the chromoelectric-dipole vertex \eqref{eq:LpNRQCDE1};
the expression for the decay rate is given in Eqs.~\eqref{eq:Gamma_spincons} and 
\eqref{eq:matrix_SO}.
The spin of the $Q\bar{Q}$ pair is the same in the initial hybrid and the final quarkonium states:
spin-$0$ hybrids decay to spin-$0$ quarkonia and spin-$1$ hybrids decay to spin-$1$ quarkonia.
For several charmonium and bottomonium spin-$0$ hybrid states, members of the hybrid multiplets $H_1\left[1^{--}\right]$, 
$H_2\left[1^{++}\right]$, $H_3\left[0^{++}\right]$ and 
their excitations, the values of the spin-conserving decay rates are shown in Table~\ref{tab:Gamma_Exclusive_spin-conserving}.
The  spin-conserving decay rates of the spin-$1$ hybrid states, members of the hybrid multiplets $H_1\left[\left(0, 1, 2\right)^{-+}\right]$, 
$H_2\left[\left(0, 1, 2\right)^{+-}\right]$, $H_3\left[1^{+-}\right]$ and their excitations, 
are, at the precision we are working (LO in the nonrelativistic expansion), three times the corresponding spin-conserving decay rates of the spin-$0$ hybrid states as the final state may assume three different polarizations, see Appendix~\ref{app:spin-matrix}.
%%%%%%%%%%%%%%%%%%%%%%%%%%%%%%%%%%%555
\begin{table}[h!]
\begin{center}
\small{\renewcommand{\arraystretch}{1.1}
\scriptsize
\begin{tabular}{|c||c|c|}
\hline
$H_m\left[J^{PC}\,\right]\,\left(\mathrm{Mass}\right)\longrightarrow Q_n \left[J^{PC}\,\right]$  &   $|\langle Q_n| \bm{r} |H_m\rangle|$~(GeV$^{-1}$)  & $\begin{array}{c} \Delta E\\\,(\text{GeV})\end{array}$ \\
\hline
\multicolumn{3}{|c|} {\hspace{0.3 cm} Charmonium hybrid}\\
\hline
$H_2\left[\,1^{++}\,\right]\,\left(4667\right) \longrightarrow \eta_c\left(1S\right)\left[\,0^{-+}\,\right]$    & 0.500 & 1.599 \\

$H_2\left[\,1^{++}\,\right]\,\left(5035\right) \longrightarrow \eta_c\left(1S\right)\left[\,0^{-+}\,\right]$    &  0.262 & 1.967\\

$H_2\left[\,1^{++}\,\right]\,\left(5035\right) \longrightarrow \eta_c\left(2S\right)\left[\,0^{-+}\,\right]$    &  0.506 & 1.358\\
\hline

$H_3\left[\,0^{++}\,\right]\,\left(5054\right) \longrightarrow \eta_c\left(1S\right)\left[\,0^{-+}\,\right]$  &  0.310 & 1.986\\

$H_3\left[\,0^{++}\,\right]\,\left(5473\right) \longrightarrow \eta_c\left(1S\right)\left[\,0^{-+}\,\right]$   & 0.150 & 2.405\\

$H_3\left[\,0^{++}\,\right]\,\left(5473\right) \longrightarrow \eta_c\left(2S\right)\left[\,0^{-+}\,\right]$  & 0.270 & 1.795\\
\hline
\multicolumn{3}{|c|} {\hspace{0.3 cm} Bottomonium hybrid}\\
\hline

$H_1\left[\,1^{--}\,\right]\,\left(10976\right) \longrightarrow h_b\left(1P\right)\left[\,1^{+-}\,\right]$   & 0.393 & 1.068 \\

$H_1\left[\,1^{--}\,\right]\,\left(11172\right) \longrightarrow h_b\left(2P\right)\left[\,1^{+-}\,\right]$   & 0.594 & 0.907 \\

\hline
$H_2\left[\,1^{++}\,\right]\,\left(10846\right) \longrightarrow \eta_b\left(1S\right)\left[\,0^{-+}\,\right]$   & 0.393 & 1.404\\

$H_2\left[\,1^{++}\,\right]\,\left(11060\right) \longrightarrow \eta_b\left(1S\right)\left[\,0^{-+}\,\right]$    & 0.321 & 1.617\\

$H_2\left[\,1^{++}\,\right]\,\left(11060\right) \longrightarrow \eta_b\left(2S\right)\left[\,0^{-+}\,\right]$   & 0.049 & 1.050\\

$H_2\left[\,1^{++}\,\right]\,\left(11270\right) \longrightarrow \eta_b\left(1S\right)\left[\,0^{-+}\,\right]$    & 0.240 & 1.828\\

$H_2\left[\,1^{++}\,\right]\,\left(11270\right) \longrightarrow \eta_b\left(2S\right)\left[\,0^{-+}\,\right]$   & 0.196 & 1.261\\

$H_2\left[\,1^{++}\,\right]\,\left(11270\right) \longrightarrow \eta_b\left(3S\right)\left[\,0^{-+}\,\right]$   & 0.214 & 0.914\\
\hline
$H_3\left[\,0^{++}\,\right]\,\left(11065\right) \longrightarrow \eta_b\left(1S\right)\left[\,0^{-+}\,\right]$  &  0.497 & 1.622 \\

$H_3\left[\,0^{++}\,\right]\,\left(11352\right) \longrightarrow \eta_b\left(1S\right)\left[\,0^{-+}\,\right]$  &   0.284 & 1.909\\

$H_3\left[\,0^{++}\,\right]\,\left(11352\right) \longrightarrow \eta_b\left(2S\right)\left[\,0^{-+}\,\right]$  &  0.499 & 1.342 \\

$H_3\left[\,0^{++}\,\right]\,\left(11616\right) \longrightarrow \eta_b\left(1S\right)\left[\,0^{-+}\,\right]$   & 0.179 & 2.174\\

$H_3\left[\,0^{++}\,\right]\,\left(11616\right) \longrightarrow \eta_b\left(2S\right)\left[\,0^{-+}\,\right]$   & 0.270 & 1.607\\
\hline
\end{tabular}
\caption{The values of the matrix element $|\langle Q_n| \bm{r} |H_m\rangle|$ and the energy difference $\Delta E=E_m^{Q\bar{Q}g}-E_{n}^{Q\bar{Q}}$
for the transition widths computed in Table~\ref{tab:Gamma_Exclusive_spin-conserving}.}
\label{tab:rDeltaE}
}
\end{center}
\end{table}

In Table~\ref{tab:Gamma_Exclusive_spin-conserving}, we list the spin-conserving transitions to below threshold quarkonia for which the decay rates can be reliably estimated in weakly-coupled pNRQCD.
These are the transitions that satisfy the conditions \eqref{eq:DELambda} and \eqref{eq:DEr}.
In practice, we require: 
$\Delta E \equiv  E_m^{Q\bar{Q}g}-E_n^{Q\bar{Q}}\gtrsim 0.8$~GeV, $\alpha_{\rm s}(\Delta E) \lesssim 0.4$ 
and $|\langle Q_n| \bm{r} |H_m\rangle| \, \Delta E \lesssim 0.8$ 
(for this last condition see Table~\ref{tab:rDeltaE}).
The  strong coupling $\alpha_{\rm s}(\Delta E)$ is evaluated at the scale $\Delta E$ 
with one-loop running~\cite{Chetyrkin:2000yt}.
We note that for the hybrid states $H_2[1^{++}]$ it holds that $L=L_{Q\bar{Q}}$;
in the charmonium case the width to $\eta_c(1S)$ may be as large as 65~MeV for 
$H_2[1^{++}](4667)$ and in the bottomonium case the width to $\eta_b(1S)$ may be as large as 29~MeV for $H_2[1^{++}](10846)$.

The spin-flipping decays of hybrids to below threshold quarkonia, $H_m \rightarrow Q_n+X$, where $X$ denotes light hadrons, are induced by the chromomagnetic-dipole vertex \eqref{eq:LpNRQCDM1};
the expression for the decay rate is given in Eqs.~\eqref{eq:Gamma_spincons} and \eqref{eq:matrix_SO-spin}.
The spin of the heavy quark-antiquark pair $\left(Q\bar {Q}\right)$ is different in the initial state hybrid and the final state quarkonium:
spin-$0$ hybrids decay to spin-$1$ quarkonia and spin-$1$ hybrids decay to spin-$0$ quarkonia.
For several charmonium and bottomonium spin-$0$ hybrid states, members of the hybrid multiplets $H_1\left[1^{--}\right]$, $H_2\left[1^{++}\right]$, $H_3\left[0^{++}\right]$,  $H_4\left[2^{++}\right]$ and their excitations, the values of the spin-flipping decay rates are shown in Table~\ref{tab:Gamma_exclusive_spin-flipping}.
The  spin-flipping decay rates of the spin-$1$ hybrid states, members of the  hybrid multiplets $H_1\left[\left(0, 1, 2\right)^{-+}\right]$,
$H_2\left[\left(0, 1, 2\right)^{+-}\right]$, $H_3\left[1^{+-}\right]$, $H_4\left[\left(1, 2, 3\right)^{+-}\right]$ and their excitations, are, at LO in the nonrelativistic expansion,  
$1/3$ the corresponding spin-flipping decay rates of the spin-$0$ hybrid states.

In Table~\ref{tab:Gamma_exclusive_spin-flipping}, we list spin-flipping transitions to below threshold quarkonia for which the decay rates can be reliably estimated in weakly-coupled pNRQCD.  
These are the transitions that satisfy the condition \eqref{eq:DELambda}.
In practice, we require $\Delta E \gtrsim 0.8$~GeV and $\alpha_{\rm s}(\Delta E) \lesssim 0.4$.
We note that, although there is no obvious hierarchy between spin-conserving and spin-flipping transition widths,
nevertheless the spin-flipping transitions tend to be smaller 
than the spin-conserving ones. This is particularly true for bottomonium hybrids.

\begin{table}[t!]
\begin{center}
\small{\renewcommand{\arraystretch}{1.0}
\scriptsize
\begin{minipage}{.49\linewidth}
\begin{tabular}{|c||c|c||c||c|}
\hline
$H_m\left[J^{PC}\,\right]\,\left(\mathrm{Mass}\right)\longrightarrow Q_n \left[J^{PC}\,\right]$ &  $\begin{array}{c}\Gamma \,\,({\rm MeV})\end{array}$\\
\hline
\multicolumn{2}{|c|} {Charmonium hybrid decay}\\
\hline
$H_1\left[\,1^{--}\,\right]\,\left(4155\right) \longrightarrow J/\psi\left(1S\right)\left[\,1^{--}\,\right]$    & 104 $^{+55}_{-26}$ $^{+49}_{-37}$   \\

$H_1\left[\,1^{--}\,\right]\,\left(4507\right) \longrightarrow J/\psi\left(1S\right)\left[\,1^{--}\,\right]$   & 46 $^{+20}_{-10}$ $^{+16}_{-13}$ \\

$H_1\left[\,1^{--}\,\right]\,\left(4507\right) \longrightarrow J/\psi\left(2S\right)\left[\,1^{--}\,\right]$   & 29 $^{+19}_{-8}$ $^{+19}_{-13}$ \\

$H_1\left[\,1^{--}\,\right]\,\left(4812\right) \longrightarrow J/\psi\left(1S\right)\left[\,1^{--}\,\right]$   & 0.6 $^{+0.2}_{-0.1}$ $^{+0.2}_{-0.1}$ \\

$H_1\left[\,1^{--}\,\right]\,\left(4812\right) \longrightarrow J/\psi\left(2S\right)\left[\,1^{--}\,\right]$   & 20 $^{+10}_{-5}$ $^{+9}_{-7}$ \\

\hline
$H_2\left[\,1^{++}\,\right]\,\left(4286\right) \longrightarrow \chi_c\left(1P\right)\left[\,\left(0, 1, 2\right)^{++}\,\right]$   & 55 $^{+38}_{-16}$ $^{+38}_{-26}$ \\

$H_2\left[\,1^{++}\,\right]\,\left(4667\right) \longrightarrow \chi_c\left(1P\right)\left[\,\left(0, 1, 2\right)^{++}\,\right]$   & 15 $^{+8}_{-4}$ $^{+7}_{-5}$ \\

$H_2\left[\,1^{++}\,\right]\,\left(5035\right) \longrightarrow \chi_c\left(1P\right)\left[\,\left(0, 1, 2\right)^{++}\,\right]$   & 6 $^{+3}_{-1}$ $^{+2}_{-2}$ \\

%$H_2\left[\,1^{++}\,\right]\,\left(5035\right) \longrightarrow \chi_c\left(1P\right)\left[\,\left(0, 1, 2\right)^{++}\,\right]$   & 11 $^{+6}_{-3}$ $^{+5}_{-4}$ \\
\hline
$H_3\left[\,0^{++}\,\right]\,\left(4590\right) \longrightarrow \chi_c\left(1P\right)\left[\,\left(0, 1, 2\right)^{++}\,\right]$    & 137 $^{+72}_{-34}$  $^{+64}_{-49}$ \\

$H_3\left[\,0^{++}\,\right]\,\left(5054\right) \longrightarrow \chi_c\left(1P\right)\left[\,\left(0, 1, 2\right)^{++}\,\right]$     & 5 $^{+2}_{-1}$ $^{+2}_{-1}$  \\

$H_3\left[\,0^{++}\,\right]\,\left(5473\right) \longrightarrow \chi_c\left(1P\right)\left[\,\left(0, 1, 2\right)^{++}\,\right]$     & 2 $^{+1}_{-0.4}$ $^{+1}_{-0.5}$  \\

\hline
$H_4\left[\,2^{++}\,\right]\,\left(4367\right) \longrightarrow \chi_c\left(1P\right)\left[\,\left(0, 1, 2\right)^{++}\,\right]$    & 65 $^{+41}_{-18}$ $^{+40}_{-28}$   \\
\hline
\multicolumn{2}{|c|} {Bottomonium hybrid decay}\\
\hline
$H_1\left[\,1^{--}\,\right[\,\left(10786\right) \longrightarrow \Upsilon\left(1S\right)\left[\,1^{--}\,\right]$     & 9 $^{+4}_{-2}$ $^{+3}_{-3}$  \\

$H_1\left[\,1^{--}\,\right]\,\left(10976\right) \longrightarrow \Upsilon\left(1S\right)\left[\,1^{--}\,\right]$    & 8 $^{+3}_{-2}$ $^{+3}_{-2}$ \\

$H_1\left[\,1^{--}\,\right]\,\left(10976\right) \longrightarrow \Upsilon\left(2S\right)\left[\,1^{--}\,\right]$   & 0.3 $^{+0.2}_{-0.1}$ $^{+0.2}_{-0.1}$  \\

$H_1\left[\,1^{--}\,\right]\,\left(11172\right) \longrightarrow \Upsilon\left(1S\right)\left[\,1^{--}\,\right]$   & 3 $^{+1}_{-1}$ $^{+1}_{-1}$    \\

$H_1\left[\,1^{--}\,\right]\,\left(11172\right) \longrightarrow \Upsilon\left(2S\right)\left[\,1^{--}\,\right]$     & 0.3 $^{+0.1}_{-0.1}$  $^{+0.1}_{-0.1}$  \\

$H_1\left[\,1^{--}\,\right]\,\left(11172\right) \longrightarrow \Upsilon\left(3S\right)\left[\,1^{--}\,\right]$     & 0.4 $^{+0.3}_{-0.1}$  $^{+0.2}_{-0.2}$  \\

\hline
$H_2\left[\,1^{++}\,\right]\,\left(10846\right) \longrightarrow \chi_b\left(1P\right)\left[\,\left(0, 1, 2\right)^{++}\,\right]$    & 6 $^{+3}_{-1}$ $^{+3}_{-2}$  \\

$H_2\left[\,1^{++}\,\right]\,\left(11060\right) \longrightarrow \chi_b\left(1P\right)\left[\,\left(0, 1, 2\right)^{++}\,\right]$     & 3 $^{+2}_{-1}$ $^{+1}_{-1}$  \\

$H_2\left[\,1^{++}\,\right]\,\left(11060\right) \longrightarrow \chi_b\left(2P\right)\left[\,\left(0, 1, 2\right)^{++}\,\right]$     & 2 $^{+1}_{-0.5}$ $^{+1}_{-1}$  \\

$H_2\left[\,1^{++}\,\right]\,\left(11270\right) \longrightarrow \chi_b\left(1P\right)\left[\,\left(0, 1, 2\right)^{++}\,\right]$     & 2 $^{+1}_{-0.4}$ $^{+1}_{-1}$  \\

$H_2\left[\,1^{++}\,\right]\,\left(11270\right) \longrightarrow \chi_b\left(2P\right)\left[\,\left(0, 1, 2\right)^{++}\,\right]$     & 2 $^{+1}_{-1}$ $^{+1}_{-1}$  \\
\hline
$H_3\left[\,0^{++}\,\right]\,\left(11065\right) \longrightarrow \chi_b\left(1P\right)\left[\,\left(0, 1, 2\right)^{++}\,\right]$     & 13 $^{+6}_{-3}$ $^{+6}_{-4}$\\

$H_3\left[\,0^{++}\,\right]\,\left(11352\right) \longrightarrow \chi_b\left(1P\right)\left[\,\left(0, 1, 2\right)^{++}\,\right]$   & 2 $^{+1}_{-1}$ $^{+1}_{-1}$   \\

$H_3\left[\,0^{++}\,\right]\,\left(11352\right) \longrightarrow \chi_b\left(2P\right)\left[\,\left(0, 1, 2\right)^{++}\,\right]$      & 9 $^{+5}_{-2}$ $^{+4}_{-3}$\\

$H_3\left[\,0^{++}\,\right]\,\left(11616\right) \longrightarrow \chi_b\left(1P\right)\left[\,\left(0, 1, 2\right)^{++}\,\right]$     & 1 $^{+0.4}_{-0.2}$ $^{+0.3}_{-0.2}$   \\

$H_3\left[\,0^{++}\,\right]\,\left(11616\right) \longrightarrow \chi_b\left(2P\right)\left[\,\left(0, 1, 2\right)^{++}\,\right]$     & 2 $^{+1}_{-0.4}$ $^{+1}_{-1}$   \\

$H_3\left[\,0^{++}\,\right]\,\left(11616\right) \longrightarrow \chi_b\left(3P\right)\left[\,\left(0, 1, 2\right)^{++}\,\right]$     & 9 $^{+5}_{-2}$ $^{+4}_{-3}$   \\
\hline
\end{tabular}
\end{minipage}
%---------------------------------------
\begin{minipage}{.49\linewidth}
\begin{tabular}{|c||c|c||c||c|}
\hline
$H_m\left[J^{PC}\,\right]\,\left(\mathrm{Mass}\right)\longrightarrow Q_n \left[J^{PC}\,\right]$ &  $\begin{array}{c}\Gamma \,\,({\rm MeV})\end{array}$\\
\hline
\multicolumn{2}{|c|} {Charmonium hybrid decay}\\
\hline

$H_1\left[\,\left(0, 1, 2\right)^{-+}\,\right]\,\left(4155\right) \longrightarrow \eta_c\left(1S\right)\left[\,0^{-+}\,\right]$    & 35 $^{+18}_{-9}$ $^{+16}_{-12}$   \\

$H_1\left[\,\left(0, 1, 2\right)^{-+}\,\right]\,\left(4507\right) \longrightarrow \eta_c\left(1S\right)\left[\,0^{-+}\,\right]$  &  15 $^{+7}_{-3}$ $^{+5}_{-4}$\\

$H_1\left[\,\left(0, 1, 2\right)^{-+}\,\right]\,\left(4507\right) \longrightarrow \eta_c\left(2S\right)\left[\,0^{-+}\,\right]$  &  10 $^{+6}_{-3}$ $^{+6}_{-4}$\\

$H_1\left[\,\left(0, 1, 2\right)^{-+}\,\right]\,\left(4812\right) \longrightarrow \eta_c\left(1S\right)\left[\,0^{-+}\,\right]$  &  0.2 $^{+0.1}_{-0.04}$ $^{+0.1}_{-0.05}$\\

$H_1\left[\,\left(0, 1, 2\right)^{-+}\,\right]\,\left(4812\right) \longrightarrow \eta_c\left(2S\right)\left[\,0^{-+}\,\right]$  &  7 $^{+3}_{-2}$ $^{+3}_{-2}$\\

\hline

$H_2\left[\,\left(0, 1, 2\right)^{+-}\,\right]\,\left(4286\right) \longrightarrow h_c\left(1P\right)\left[\,1^{+-}\,\right]$   & 18 $^{+13}_{-5}$ $^{+13}_{-9}$
\\

$H_2\left[\,\left(0, 1, 2\right)^{+-}\,\right]\,\left(4667\right) \longrightarrow h_c\left(1P\right)\left[\,1^{+-}\,\right]$   & 5 $^{+3}_{-1}$ $^{+2}_{-2}$
\\

$H_2\left[\,\left(0, 1, 2\right)^{+-}\,\right]\,\left(5035\right) \longrightarrow h_c\left(1P\right)\left[\,1^{+-}\,\right]$   & 2 $^{+1}_{-0.4}$ $^{+1}_{-1}$
\\

\hline
$H_3\left[\,1^{+-}\,\right]\,\left(4590\right) \longrightarrow h_c\left(1P\right)\left[\,1^{+-}\,\right]$    &  46 $^{+24}_{-11}$ $^{+21}_{-16}$ \\

$H_3\left[\,1^{+-}\,\right]\,\left(5054\right) \longrightarrow h_c\left(1P\right)\left[\,1^{+-}\,\right]$     & 2 $^{+1}_{-0.4}$ $^{+1}_{-0.4}$  \\

$H_3\left[\,1^{+-}\,\right]\,\left(5473\right) \longrightarrow h_c\left(1P\right)\left[\,1^{+-}\,\right]$     & 0.7 $^{+0.3}_{-0.1}$ $^{+0.2}_{-0.2}$  \\
\hline

$H_4\left[\,\left(1, 2, 3\right)^{+-}\,\right]\,\left(4367\right) \longrightarrow h_c\left(1P\right)\left[\,1^{+-}\,\right]$    & 22 $^{+14}_{-6}$ $^{+13}_{-9}$   \\

\hline
\multicolumn{2}{|c|} {Bottomonium hybrid decay}\\
\hline
$H_1\left[\,\left(0, 1, 2\right)^{-+}\,\right]\,\left(10786\right) \longrightarrow \eta_b\left(1S\right)\left[\,0^{-+}\,\right]$     & 3 $^{+1}_{-1}$ $^{+1}_{-1}$   \\

$H_1\left[\,\left(0, 1, 2\right)^{-+}\,\right]\,\left(10976\right) \longrightarrow \eta_b\left(1S\right)\left[\,0^{-+}\,\right]$    & 3 $^{+1}_{-1}$ $^{+1}_{-1}$  \\

$H_1\left[\,\left(0, 1, 2\right)^{-+}\,\right]\,\left(10976\right) \longrightarrow \eta_b\left(2S\right)\left[\,0^{-+}\,\right]$   & 0.1 $^{+0.1}_{-0.02}$ $^{+0.1}_{-0.04}$   \\

$H_1\left[\,\left(0, 1, 2\right)^{-+}\,\right]\,\left(11172\right) \longrightarrow \eta_b\left(1S\right)\left[\,0^{-+}\,\right]$   & 1 $^{+0.4}_{-0.2}$ $^{+0.3}_{-0.3}$    \\

$H_1\left[\,\left(0, 1, 2\right)^{-+}\,\right]\,\left(11172\right) \longrightarrow \eta_b\left(2S\right)\left[\,0^{-+}\,\right]$     & 0.1 $^{+0.04}_{-0.02}$ $^{+0.04}_{-0.03}$ \\

$H_1\left[\,\left(0, 1, 2\right)^{-+}\,\right]\,\left(11172\right) \longrightarrow \eta_b\left(3S\right)\left[\,0^{-+}\,\right]$     & 0.1 $^{+0.08}_{-0.04}$ $^{+0.08}_{-0.06}$ \\

\hline
$H_2\left[\,\left(0, 1, 2\right)^{+-}\,\right]\,\left(10846\right) \longrightarrow h_b\left(1P\right)\left[\,1^{+-}\,\right]$    & 2 $^{+1}_{-0.5}$ $^{+1}_{-1}$  \\

$H_2\left[\,\left(0, 1, 2\right)^{+-}\,\right]\,\left(11060\right) \longrightarrow h_b\left(1P\right)\left[\,1^{+-}\,\right]$     & 1 $^{+1}_{-0.3}$ $^{+0.5}_{-0.4}$ \\

$H_2\left[\,\left(0, 1, 2\right)^{+-}\,\right]\,\left(11060\right) \longrightarrow h_b\left(2P\right)\left[\,1^{+-}\,\right]$     & 0.5 $^{+0.4}_{-0.2}$ $^{+0.4}_{-0.3}$ \\

$H_2\left[\,\left(0, 1, 2\right)^{+-}\,\right]\,\left(11270\right) \longrightarrow h_b\left(1P\right)\left[\,1^{+-}\,\right]$     & 1 $^{+0.3}_{-0.1}$ $^{+0.2}_{-0.2}$ \\

$H_2\left[\,\left(0, 1, 2\right)^{+-}\,\right]\,\left(11270\right) \longrightarrow h_b\left(2P\right)\left[\,1^{+-}\,\right]$     & 1 $^{+0.4}_{-0.2}$ $^{+0.3}_{-0.3}$ \\
\hline
$H_3\left[\,1^{+-}\,\right]\,\left(11065\right) \longrightarrow h_b\left(1P\right)\left[\,1^{+-}\,\right]$     & 4 $^{+2}_{-1}$ $^{+2}_{-1}$ \\

$H_3\left[\,1^{+-}\,\right]\,\left(11352\right) \longrightarrow h_b\left(1P\right)\left[\,1^{+-}\,\right]$     & 1 $^{+0.4}_{-0.2}$ $^{+0.3}_{-0.2}$ \\

$H_3\left[\,1^{+-}\,\right]\,\left(11352\right) \longrightarrow h_b\left(2P\right)\left[\,1^{+-}\,\right]$     & 3 $^{+2}_{-1}$ $^{+1}_{-1}$  \\

$H_3\left[\,1^{+-}\,\right]\,\left(11616\right) \longrightarrow h_b\left(1P\right)\left[\,1^{+-}\,\right]$     & 0.3 $^{+0.1}_{-0.1}$ $^{+0.1}_{-0.1}$ \\

$H_3\left[\,1^{+-}\,\right]\,\left(11616\right) \longrightarrow h_b\left(2P\right)\left[\,1^{+-}\,\right]$     & 1 $^{+0.3}_{-0.1}$ $^{+0.2}_{-0.2}$ \\

$H_3\left[\,1^{+-}\,\right]\,\left(11616\right) \longrightarrow h_b\left(3P\right)\left[\,1^{+-}\,\right]$     & 3 $^{+2}_{-1}$ $^{+1}_{-1}$  \\
\hline
\end{tabular}
\end{minipage}
\caption{Spin-flipping semi-inclusive decay rates of hybrids decaying to quarkonia below threshold, 
due to the chromomagnetic-dipole interaction~\eqref{eq:LpNRQCDM1}.
The decay rates are computed from Eqs.~\eqref{eq:Gamma_spincons} and \eqref{eq:matrix_SO-spin}.
The hybrid states are denoted by $H_m\left[J^{PC}\,\right]\,\left(\mathrm{mass}\right)$, 
where $J$ is the total angular momentum quantum number including the spin, and the masses are in MeV.
The quarkonium states are denoted by the physical states. 
For the quarkonium states, we use the spin-averaged masses given in Table~\ref{tab:QQbarspectrum}.
The first error comes from varying the scale of $\alpha_{\rm s}$ from $\Delta E/2$ to $2\,\Delta E$, and the second one from the gluelump mass uncertainty of $\pm\, 0.15\,\mathrm{GeV}$.
We only show decay rates for which $\Delta E \gtrsim 0.8$~GeV  
and $\alpha_{\rm s}(\Delta E) \lesssim 0.4$.}
\label{tab:Gamma_exclusive_spin-flipping}
}
\end{center}
\end{table}
\normalsize

\begin{figure}[h!]
\begin{center}
\includegraphics[width=0.70\textwidth,angle=0,clip]{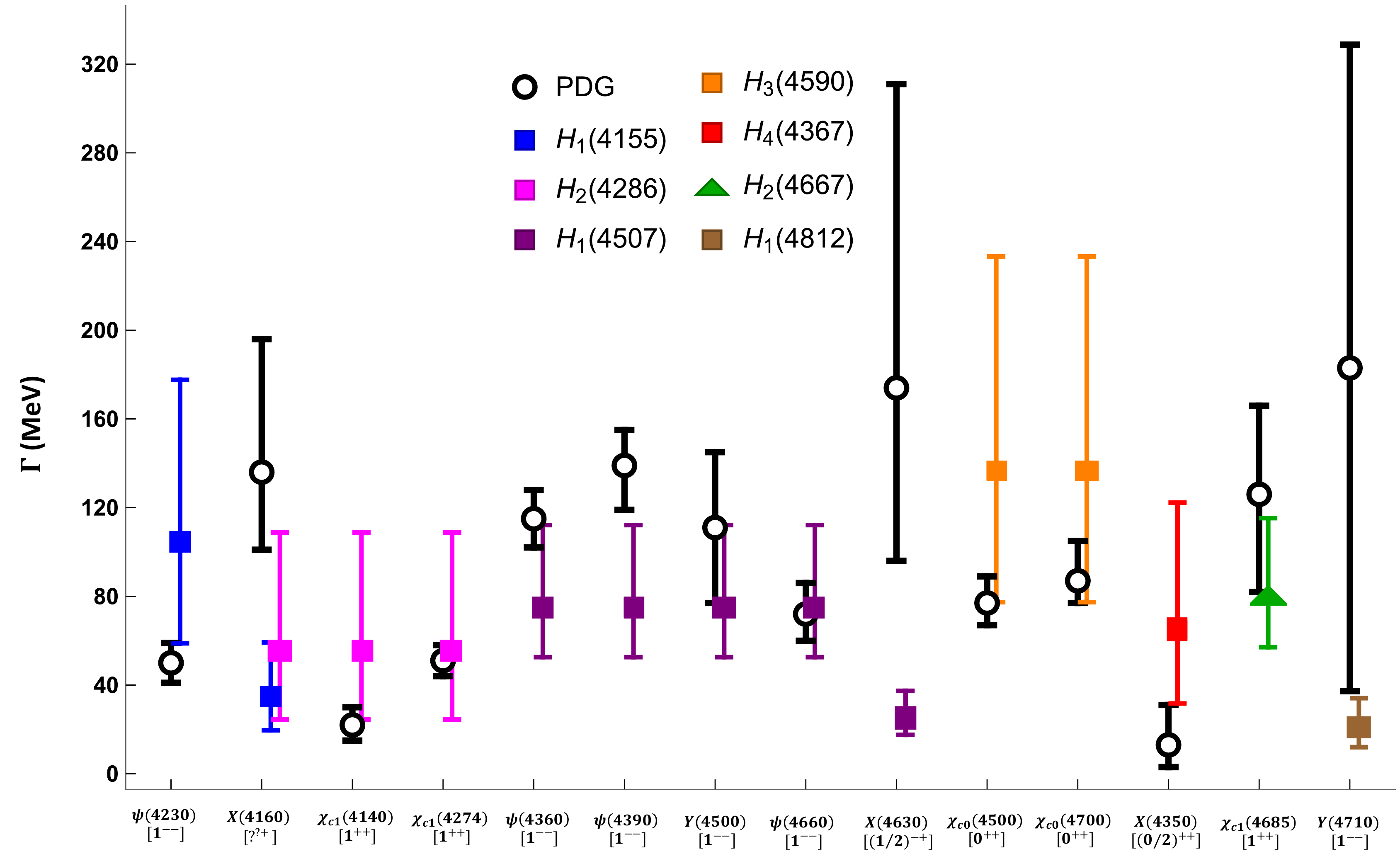}
\caption{Comparison of the total decay widths of the neutral exotic charmonium states from  Table~\ref{tab:exotic} with the hybrid-to-quarkonium transition widths 
computed in this work according to the assignments in Fig.~\ref{fig:cchybrids}.
For $H_2 \left(4667\right)$ (represented by triangles), the transition width is the sum of the 
spin-conserving transition width in Table~\ref{tab:Gamma_Exclusive_spin-conserving} and the 
spin-flipping transition width in Table~\ref{tab:Gamma_exclusive_spin-flipping}.
For all other hybrid states (represented by squares), the transition widths 
are just given by the spin-flipping transition widths in Table~\ref{tab:Gamma_exclusive_spin-flipping} 
as the spin-conserving transitions violate the condition \eqref{eq:DEr}.}
\label{fig:decay_rate_charm}
\end{center}
\end{figure}

\begin{figure}[h]
\begin{center}
\includegraphics[width=0.65\textwidth,angle=0,clip]{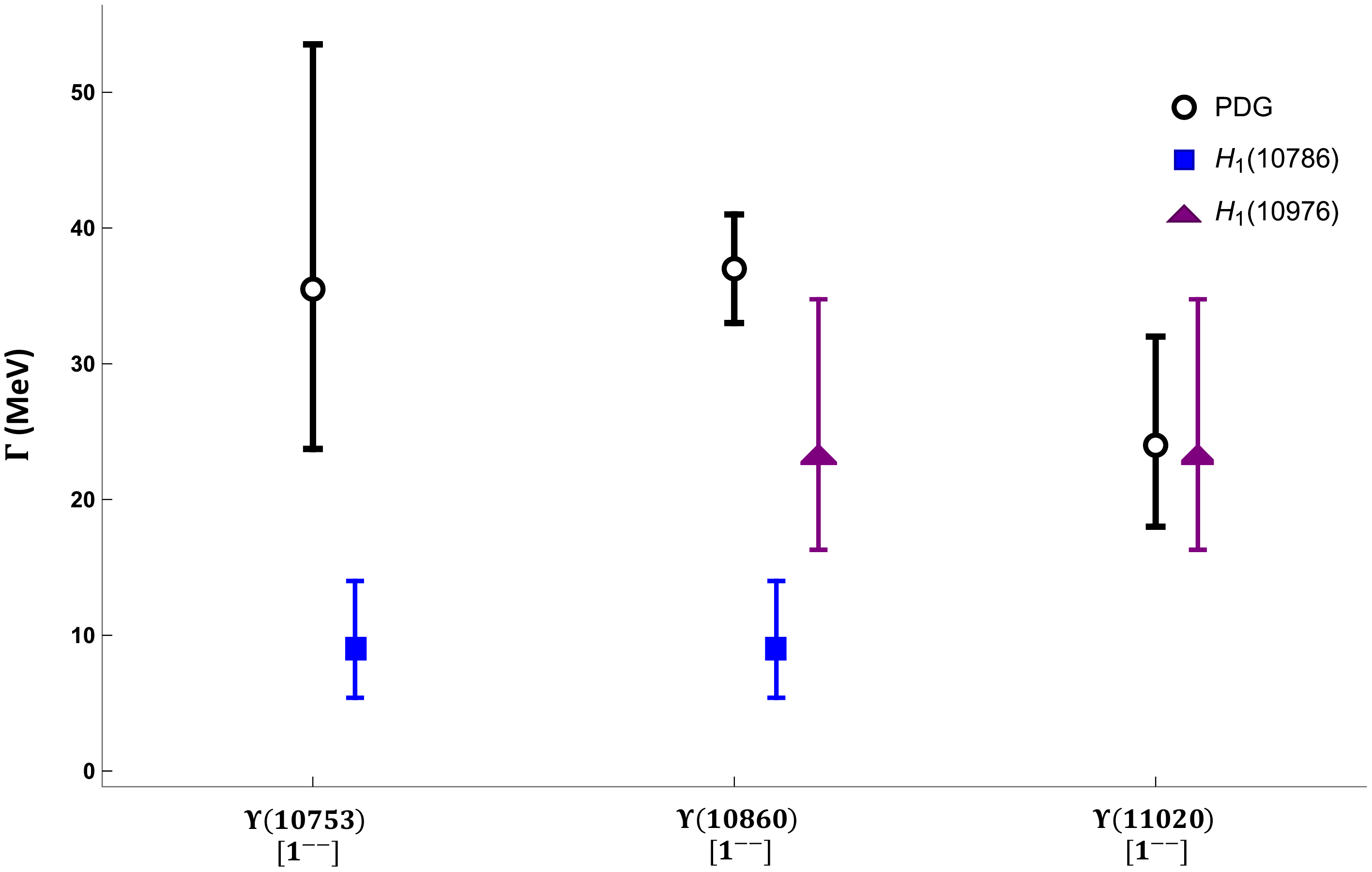}
\caption{
  Comparison of the total decay widths of the neutral exotic bottomonium states from Table~\ref{tab:exotic}
  with the hybrid-to-quarkonium transition widths computed in this work according to the assignments in Fig.~\ref{fig:bbhybrids}.
For $H_1 \left(10976\right)$ (represented by triangles), the transition width is the sum of the 
spin-conserving transition width in Table~\ref{tab:Gamma_Exclusive_spin-conserving} and the 
spin-flipping transition widths in Table~\ref{tab:Gamma_exclusive_spin-flipping}.
For all other hybrid states (represented by squares), the transition widths 
are just given by the spin-flipping transition widths in Table~\ref{tab:Gamma_exclusive_spin-flipping} 
as the spin-conserving transitions violate the condition \eqref{eq:DEr}.}
\label{fig:decay_rate_bottom}
\end{center}
\end{figure}

In Figs.~\ref{fig:decay_rate_charm} and \ref{fig:decay_rate_bottom} we compare 
the measured total decay widths of the neutral exotic charmonium states from Table~\ref{tab:exotic} with the hybrid-to-quarkonium transition widths 
computed in this work and listed in the Tables~\ref{tab:Gamma_Exclusive_spin-conserving} 
and~\ref{tab:Gamma_exclusive_spin-flipping},  according to the assignments made in Figs.~\ref{fig:cchybrids} and~\ref{fig:bbhybrids}.
The total decay width is the sum of all exclusive decay widths, 
therefore the hybrid-to-quarkonium transition widths computed in this work can 
only provide a lower bound for the hybrid total decay width.
Moreover, for the charmonium hybrid states $H_1\left(4155\right)$, $H_1\left(4507\right)$, $H_1\left(4812\right)$, $H_2\left(4286\right)$, $H_3\left(4590\right)$, $H_4\left(4367\right)$
and the bottomonium hybrid state $H_1\left(10786\right)$ 
we cannot reliably estimate the spin-conserving transition widths due to violation of the condition \eqref{eq:DEr}.
Hence, for these states we show in Figs.~\ref{fig:decay_rate_charm} and \ref{fig:decay_rate_bottom} 
only the sum of the spin-flipping transition widths listed in Table~\ref{tab:Gamma_exclusive_spin-flipping}.
For the charmonium hybrid $H_2\left(4667\right)$ and the bottomonium hybrid $H_1\left(10976\right)$, both the spin-conserving and spin-flipping transition widths could be computed 
(see Tables~\ref{tab:Gamma_Exclusive_spin-conserving} and \ref{tab:Gamma_exclusive_spin-flipping}) 
and their sum is shown in Figs.~\ref{fig:decay_rate_charm} and \ref{fig:decay_rate_bottom}.
Based on Figs.~\ref{fig:decay_rate_charm} and \ref{fig:decay_rate_bottom}, we can make 
the following observations for each state.\footnote{
We have computed masses and transition widths   
assuming that the states are either pure quarkonium or pure hybrid states. 
We are aware, however, that mixing between quarkonium and hybrid states may influence the phenomenology of the physical states~\cite{Oncala:2017hop}, 
eventually affecting some of their interpretations}.

\begin{itemize}
\item $\psi\left(4230\right)$ (also known as $Y\left(4260\right)$): The mass and quantum numbers of this state are compatible with the hybrid state $H_1\left[1^{--}\right]\left(4155\right)$ within uncertainties.
The experimental determination of the inclusive decay width of $\psi\left(4230\right)$ is $50\pm9$~MeV~\cite{Workman:2022ynf}.
Our estimate for the lower bound on the total decay width of  $H_1\left[1^{--}\right]\left(4155\right)$ is $104^{+74}_{-45}$~MeV, which is almost twice the experimental value. 
This disfavours the interpretation of $\psi\left(4230\right)$ as a pure hybrid state.
It should be mentioned, however, that our estimate could be consistent within errors with the recent measure of $73\pm32$~MeV for the inclusive decay width of $\psi\left(4230\right)$ by the BESIII experiment~\cite{BESIII:2022joj}.
%----------------------------------
\item $\psi\left(4360\right)$: The mass and quantum numbers of this state are compatible with the hybrid state $H_1\left[1^{--}\right]\left(4507\right)$ within uncertainties.
The experimental determination of the inclusive decay width of $\psi\left(4360\right)$ is $115\pm13$~MeV~\cite{Workman:2022ynf}.
Our estimate for the lower bound on the total decay width of  $H_1\left[1^{--}\right]\left(4507\right)$ is $75^{+37}_{-22}$~MeV,
which is lower, although overlapping within errors, with the experimental determination. Within present uncertainties, the state could therefore have a $H_1\left[1^{--}\right]\left(4507\right)$ hybrid component.
%----------------------------------
\item $\psi\left(4390\right)$: The mass and quantum numbers of this state are compatible with the hybrid state $H_1\left[1^{--}\right]\left(4507\right)$ within uncertainties.
The experimental determination of the inclusive decay width of $\psi\left(4390\right)$ is $139^{+16}_{-20}$~MeV~\cite{Workman:2022ynf}.
Our estimate for the lower bound on the total decay width of  $H_1\left[1^{--}\right]\left(4507\right)$ is $75^{+37}_{-22}$~MeV, which is below the experimental determination.
If the state is experimentally confirmed, it could have a significant $H_1\left[1^{--}\right]\left(4507\right)$ hybrid component.
%----------------------------------
\item $Y\left(4500\right)$: The mass and quantum numbers of this state recently seen by the BESIII experiment~\cite{BESIII:2022joj} 
in a resonance structure in the $e^{+} e^{-}\rightarrow K^+K^{-} J/\psi$ 
cross section 
are compatible with the hybrid state $H_1\left[1^{--}\right]\left(4507\right)$ within uncertainties. 
The experimental determination of the inclusive decay width of $Y\left(4500\right)$ is $111\pm 34$~MeV \cite{BESIII:2022joj}.
Our estimate for the lower bound on the total decay width of  $H_1\left[1^{--}\right]\left(4507\right)$ is $75^{+37}_{-22}$~MeV, which is consistent within errors with the experimental determination.
Within present uncertainties, the state could have a  $H_1\left[1^{--}\right]\left(4507\right)$ hybrid component.
%----------------------------------
\item $\psi\left(4660\right)$: The mass and quantum numbers of this state are compatible with the hybrid state $H_1\left[1^{--}\right]\left(4507\right)$ within uncertainties.
The experimental determination of the inclusive decay width of $\psi\left(4660\right)$ is $72^{+14}_{-12}$~MeV~\cite{Workman:2022ynf}.
Our estimate for the lower bound on the total decay width of $H_1\left[1^{--}\right]\left(4507\right)$ is $75^{+37}_{-22}$~MeV, which 
overlaps within errors with the experimental determination.
Decays of $\psi\left(4660\right)$ through open flavor channels have been detected and may eventually contribute to a large portion of the decay width.
%----------------------------------
\item $Y\left(4710\right)$: The mass and quantum numbers of this state recently seen by the BESIII experiment~\cite{Ablikim:2022yav} 
in a resonance structure in the $e^{+} e^{-}\rightarrow K^0_SK^{0}_S J/\psi$ 
cross section 
are compatible with the hybrid state $H_1\left[1^{--}\right]\left(4812\right)$ within uncertainties. 
The experimental determination of the inclusive decay width of $Y\left(4710\right)$ is $183\pm 146$~MeV~\cite{Ablikim:2022yav}.
Our estimate for the lower bound on the total decay width of  $H_1\left[1^{--}\right]\left(4812\right)$ is $21^{+13}_{-9}$~MeV, which is much lower than the central value of the experimental determination; the experimental uncertainty is however large.
This suggests that $Y\left(4710\right)$ could have a significant $H_1\left[1^{--}\right]\left(4812\right)$ hybrid state component.
%----------------------------------
\item $X\left(4160\right)$: The mass and a likely positive charge conjugation 
(the assignment $J^{PC} = 2^{-+}$ is currently favoured) of this state could make it compatible with the hybrid states $H_1\left[\left(0, 1, 2\right)^{-+}\right]\left(4155\right)$ or $H_2\left[1^{++}\right]\left(4286\right)$ within uncertainties.
The experimental determination of the inclusive decay width of $X\left(4160\right)$ is $136^{+60}_{-35}$ MeV~\cite{Workman:2022ynf}.
Our estimate for the lower bound on the total decay width of $H_1\left[\left(0, 1, 2\right)^{-+}\right]\left(4155\right)$ is $35^{+24}_{-15}$~MeV
and of $H_2\left[1^{++}\right]\left(4286\right)$ is $55^{+54}_{-31}$~MeV.
The central values of both estimates are much lower than the experimental determination, which may indicate that $X\left(4160\right)$ 
has a large hybrid state component, in particular if it is the $H_1\left[\left(0, 1, 2\right)^{-+}\right]\left(4155\right)$.
%----------------------------------
\item $\chi_{c1}\left(4140\right)$: The mass and quantum numbers of this state are compatible with the hybrid state $H_2\left[1^{++}\right]\left(4286\right)$ 
within uncertainties.
The experimental determination of the inclusive decay width of $\chi_{c1}\left(4140\right)$ is $19^{+7}_{-5}$~MeV~\cite{Workman:2022ynf}. 
Our estimate for the lower bound on the total decay width of $H_2\left[1^{++}\right]\left(4286\right)$ is $55^{+54}_{-31}$~MeV.
The central value is around three times the experimental value of the total width, and only marginally compatible within errors; this disfavours  
a large hybrid component for the state.
%----------------------------------
\item $\chi_{c1}\left(4274\right)$: The mass and quantum numbers of this state are compatible with the hybrid state $H_2\left[1^{++}\right]\left(4286\right)$ 
within uncertainties.
The experimental determination of the inclusive decay width of $\chi_{c1}\left(4274\right)$ is $51\pm7$~MeV~\cite{Workman:2022ynf}.
Our estimate for the lower bound on the total decay width of $H_2\left[1^{++}\right]\left(4286\right)$ is $55^{+54}_{-31}$~MeV, which overlaps within errors with the experimental value.
%----------------------------------
\item $\chi_{c0}\left(4500\right)$: The mass and quantum numbers of this state are compatible with the hybrid state $H_3\left[0^{++}\right]\left(4590\right)$ 
within uncertainties.
The experimental determination of the inclusive decay width of $\chi_{c0}\left(4500\right)$ is $77^{+12}_{-10}$~MeV~\cite{Workman:2022ynf}.
Our estimate for the lower bound on the total decay width of $H_3\left[0^{++}\right]\left(4590\right)$ is $137^{+96}_{-60}$ MeV.
The central value is roughly twice the experimental value of the total width,
moreover it comes from the spin-flipping decay to $\chi_c(1P)$ that has 
not been observed; this disfavours a large hybrid component for the state.
%----------------------------------
\item $\chi_{c0}\left(4700\right)$: The mass and quantum numbers of this state are compatible with the hybrid state $H_3\left[0^{++}\right]\left(4590\right)$ 
within uncertainties.
The experimental determination of the inclusive decay width of $\chi_{c0}\left(4700\right)$ is $87^{+18}_{-10}$~MeV~\cite{Workman:2022ynf}.
Our estimate for the lower bound on the total decay width of $H_3\left[0^{++}\right]\left(4590\right)$ is $137^{+96}_{-60}$~MeV.
For this state, it holds what we have written for the $\chi_{c0}\left(4500\right)$ state: our lower bound has a central 
value that is larger than the central value of the measured width, 
moreover it comes from the spin-flipping decay to $\chi_c(1P)$ that has 
not been observed. A large hybrid component for the state appears therefore disfavoured.
%----------------------------------
\item $X\left(4350\right)$: The mass and assuming quantum numbers $2^{++}$ for this state 
(also $0^{++}$ is possible, see Ref.~\cite{Belle:2009rkh})
are compatible with the hybrid state $H_4\left[2^{++}\right]\left(4367\right)$ 
within uncertainties.
The experimental determination of the inclusive decay width of $X\left(4350\right)$ is $13^{+18}_{-10}$~MeV~\cite{Workman:2022ynf}.
Our estimate for the lower bound on the total decay width of $H_4\left[2^{++}\right]\left(4367\right)$ is $65^{+57}_{-33}$~MeV,
which is almost four times the experimental value of the total width.
This disfavours the interpretation of the $X\left(4350\right)$ as a pure hybrid state.
%----------------------------------
\item $\chi_{c1}\left(4685\right)$: The mass and quantum numbers of this state are compatible with the hybrid state $H_2\left[1^{++}\right]\left(4667\right)$ 
within uncertainties.
The experimental determination of the inclusive decay width of $\chi_{c1}\left(4685\right)$ is $126^{+40}_{-44}$~MeV~\cite{Workman:2022ynf}.
Our estimate for the hybrid-to-quarkonium decay width of $H_2\left[1^{++}\right]\left(4667\right)$ is $80^{+35}_{-23}$~MeV, which is  compatible with the experimental value of the total width.
This suggests that $\chi_{c1}\left(4685\right)$ could have a $H_2\left[1^{++}\right]\left(4667\right)$ hybrid state component, 
although only decays to $\phi J/\psi$ have been seen.
%----------------------------------
\item $X\left(4630\right)$: The mass and the quantum numbers $J^{PC}=\left(1\,\mbox{or}\,2\right)^{-+}$ of this state is compatible with the hybrid states $H_1\left[\left(0, 1, 2\right)^{-+}\right]\left(4507\right)$ or $H_5\left[\left(1, 2, 3\right)^{-+}\right]\left(4476\right)$ within uncertainties.
The experimental determination of the inclusive decay width of $X\left(4630\right)$ is $174^{+137}_{-78}$~MeV~\cite{Workman:2022ynf}.
Our estimate for the lower bound on the total decay width of  $H_1\left[\left(0, 1, 2\right)^{-+}\right]\left(4507\right)$ is $25^{+12}_{-7}$~MeV which is much lower than the experimental determination. This may indicate that the  $X\left(4630\right)$ state 
has a large hybrid state component, in particular if it is the $H_1\left[\left(0, 1, 2\right)^{-+}\right]\left(4507\right)$.\footnote{ For $H_5\left[\left(1, 2, 3\right)^{-+}\right]\left(4476\right)$, we cannot estimate the hybrid-to-quarkonium decay width because the spin-conserving transition widths violate the condition \eqref{eq:DEr} and the spin-flipping transitions are to $D$-wave charmonium states, which are either above the lowest $D\bar{D}$ threshold or have not been experimentally observed.}
%----------------------------------
\item $\Upsilon\left(10753\right)$: The mass and quantum numbers of this state are compatible with the hybrid state $H_1\left[1^{--}\right]\left(10786\right)$ 
within uncertainties. 
The experimental determination of the inclusive decay width of $\Upsilon\left(10753\right)$ is $36^{+18}_{-12}$~MeV~\cite{Workman:2022ynf}.
Our estimate for the width of $H_1\left[1^{--}\right]\left(10786\right)$ 
to $\Upsilon(1S)$ is $9^{+5}_{-4}$~MeV, 
which is well in agreement with the determination $9.7\pm3.8$~MeV in 
Ref.~\cite{TarrusCastella:2021pld}.
The fact that the computed width to $\Upsilon(1S)$ is much smaller than the experimental value of the total width is consistent with  
$\Upsilon\left(10753\right)$ having a large $H_1\left[1^{--}\right]\left(10786\right)$ hybrid state component.
%----------------------------------
\item $\Upsilon\left(10860\right)$: The mass and quantum numbers of this state are compatible with the hybrid states $H_1\left[1^{--}\right]\left(10786\right)$ or $H_1\left[1^{--}\right]\left(10976\right)$ within uncertainties.
The experimental determination of the inclusive decay width of $\Upsilon\left(10860\right)$ is $37\pm 4$~MeV~\cite{Workman:2022ynf}.
If we subtract from it the fraction, $76.6\%$, of decays into open 
bottom mesons, we obtain $8.8^{+2.6}_{-1.8}$~MeV.
Our estimate for the lower bound on the total decay width of $H_1\left[1^{--}\right]\left(10786\right)$ is $9^{+5}_{-4}$~MeV, which is 
in good agreement with this latter value, whereas 
our estimate for the hybrid-to-quarkonium decay width of $H_1\left[1^{--}\right]\left(10976\right)$ is $23^{+11}_{-7}$~MeV, 
which is larger than $8.8^{+2.6}_{-1.8}$~MeV.
This leaves open the possibility that 
$\Upsilon\left(10860\right)$ is made of a conventional 
$\Upsilon(5S)$ quarkonium state mixed with a significant $H_1\left[1^{--}\right]\left(10786\right)$ hybrid state component.
%----------------------------------
\item $\Upsilon\left(11020\right)$: The mass and quantum numbers of this state are compatible with the hybrid state $H_1\left[1^{--}\right]\left(10976\right)$ 
within uncertainties.
The experimental determination of the inclusive decay width of $\Upsilon\left(11020\right)$ is $24^{+8}_{-6}$~MeV~\cite{Workman:2022ynf}.
For the $H_1\left[1^{--}\right]\left(10976\right)$ state, we could compute 
the spin-conserving transition width to $h_b(1P)$, $15^{+11}_{-6}$~MeV,  
and the spin-flipping transition widths to $\Upsilon(1S)$, $8^{+4}_{-3}$~MeV,  
and $\Upsilon(2S)$, $0.3^{+0.3}_{-0.1}$~MeV.  
Our results compare well with the spin-conserving and spin-flipping transitions computed in Ref.~\cite{TarrusCastella:2021pld}, where the authors get $20\pm 9$~MeV for the transition to $h_b(1P)$,  $7.3\pm2.5$~MeV 
for the transition to $\Upsilon(1S)$ and $1.1\pm0.5$~MeV for the transition 
to $\Upsilon(2S)$.
Summing up all the three contributions, 
our estimate for the hybrid-to-quarkonium decay width of $H_1\left[1^{--}\right]\left(10976\right)$ is $23^{+11}_{-7}$~MeV, which is of the same size as the experimental value of the total width.
The latter, however, includes also decays to open bottom hadrons.
\end{itemize}

\section{Conclusions}\label{conclusion}
In this work, we have computed semi-inclusive decay rates of low-lying quarkonium hybrids, $H_m$,  
into conventional quarkonia below threshold, $Q_n$, using the Born--Oppenheimer EFT framework~\cite{Berwein:2015vca,Oncala:2017hop,Brambilla:2017uyf,Soto:2020xpm}.
We require the decay channels to satisfy the hierarchy of scales 
$1/|\langle Q_n|\bm{r}|H_m\rangle| \gg \Delta E \gg \Lambda_{\rm QCD} \gg m_Q v^2$,
where $\Delta E$ is the mass difference between the decaying hybrid and the final-state quarkonium.
The first inequality allows  multipole expanding the gluon emitted in the transition: we work at NLO in the multipole expansion.
The second inequality allows to treat the emitted gluon in weakly-coupled perturbation theory.
The last inequality permits to neglect quarkonium hybrids of higher-lying gluonic excitations,
$m_Q v^2$ being the typical energy splittings for a nonrelativistic bound state in a given potential. 
At NLO in the multipole expansion and at order $1/m_Q$ in the nonrelativistic expansion,
two hybrid-to-quarkonium decay channels are possible: a spin-conserving one induced by the chromoelectric-dipole interaction \eqref{eq:LpNRQCDE1}, whose width is given by Eqs.~\eqref{eq:Gamma_spincons} and \eqref{eq:matrix_SO}, 
and a spin-flipping one induced by the chromomagnetic-dipole interaction \eqref{eq:LpNRQCDM1},  
whose width is given by Eqs.~\eqref{eq:Gamma_spincons} and \eqref{eq:matrix_SO-spin}.
The relative size of the corresponding two decay widths is dictated by the 
energy difference between hybrid and quarkonium state, 
and by the dimensionless quantity $m_Q\,|\langle Q_n|\bm{r}|H_m\rangle|/|\langle Q_n|H_m\rangle|$, which is not necessarily large, in particular in the charmonium hybrid sector.
Spin-flipping transitions may, therefore, compete under some circumstances with spin-conserving ones. The situation is somewhat 
different from what happens in common  quarkonium-to-quarkonium transitions, where spin-conserving transitions are enhanced with respect to 
spin-flipping non-hindered (chromo)magnetic transitions
by the matrix element,  
$(m_Q\,|\langle Q_n|\bm{r}|Q_m\rangle|)^2 \sim 1/v^2 \gg 1$,   
and by the large energy gap between the initial and final state quarkonium.

The results for the hybrid-to-quarkonium decay widths are listed in the Tables~\ref{tab:Gamma_Exclusive_spin-conserving} and~\ref{tab:Gamma_exclusive_spin-flipping}.
They supersede, confirm or add to previously obtained results in a similar framework~\cite{Oncala:2017hop,TarrusCastella:2021pld}.
We may relate hybrid states with some of the XYZ states discovered in the last decades 
in the charmonium and bottomonium sector by comparing masses and quantum numbers.
This is done in Figs.~\ref{fig:cchybrids} and~\ref{fig:bbhybrids}, which update similar figures in Refs.~\cite{Berwein:2015vca,Brambilla:2019esw}.
After assigning hybrid to physical states, the hybrid-to-quarkonium widths in the Tables~\ref{tab:Gamma_Exclusive_spin-conserving}
and~\ref{tab:Gamma_exclusive_spin-flipping} provide lower bounds on the widths of the physical states, if interpreted as pure hybrid states.
The comparison of these lower bounds with the measured widths of the XYZ states is made in Figs.~\ref{fig:decay_rate_charm} and~\ref{fig:decay_rate_bottom}. 
 
Figures~\ref{fig:decay_rate_charm} and~\ref{fig:decay_rate_bottom} show that hybrid-to-quarkonium widths constrain the hybrid interpretation of the XYZ states much more strongly than just quantum numbers and masses. 
In particular our calculations disfavour the interpretation of 
$\psi(4230)$, $\chi_{c1}(4140)$, $\chi_{c0}(4500)$, $\chi_{c0}(4700)$ 
and $X(4350)$ as pure hybrid states, 
while they favour a significant hybrid component $H_1[2^{-+}]$ in $X(4160)$, 
$H_1[1^{--}]$ in $\psi(4390)$, if the state is experimentally confirmed, 
$H_1[(1,2)^{-+}]$ or $H_5[(1,2)^{-+}]$ in $X(4630)$, 
$H_1[1^{--}]$ in $Y(4710)$,
and, in the bottomonium sector, a large hybrid component 
$H_1[1^{--}]$ in $\Upsilon(10753)$ and in $\Upsilon(10860)$.
For the other states no definite conclusions can be drawn.
A more detailed discussion can be found at the end of Sec.~\ref{subsec:decay_results}.

The study presented in this work can be improved both theoretically and phenomenologically in several ways.
On the theoretical side, the framework may require a more systematic implementation of nonperturbative 
effects, responsible for the binding, and weakly-coupled effects responsible for the decay to quarkonium, 
for instance, to better justify promoting color octet and color singlet weakly-coupled Hamiltonians to hybrid and quarkonium Hamiltonians, 
or using the spectator gluon approximation to evaluate four-field correlators.
Also desirable is the enlargement of the EFT degrees of freedom to encompass open heavy-flavor states,
which may have a large impact on the physics of states above the open flavor threshold~\cite{TarrusCastella:2022rxb}.
On the phenomenological side, accounting for the mixing of hybrid and quarkonium pure states 
may have an important effect on some states, and eventually alter the interpretation of some of the XYZ exotics.
The mixing potential between hybrid and quarkonium has been constrained in the long and short range in 
Ref.~\cite{Oncala:2017hop}. Ideally it should be determined in lattice QCD, but such a computation is not available yet.

\section*{Acknowledgements}
This work has been supported by the DFG Project-ID 196253076 TRR 110 and the NSFC through
funds provided to the Sino-German CRC 110 ``Symmetries and the Emergence of Structure in QCD''.
We thank  Joan Soto and Jaume Tarr\'us Castell\`a for several useful discussions, and 
Roberto Mussa and Changzheng Yuan for communications.

\appendix

\section{Quarkonium wavefunctions}\label{app:QQbarwf}
\begin{figure}[ht]
\begin{subfigure}{.4\textwidth}
  \centering
  \includegraphics[width=0.85\linewidth]{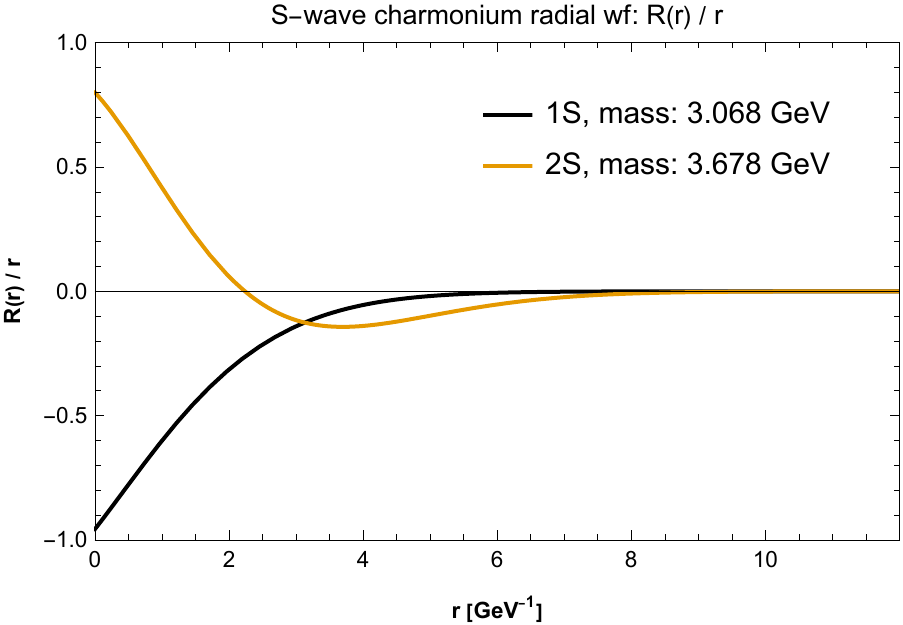}
\end{subfigure}
\begin{subfigure}{.4\textwidth}
  \centering
  \includegraphics[width=0.85\linewidth]{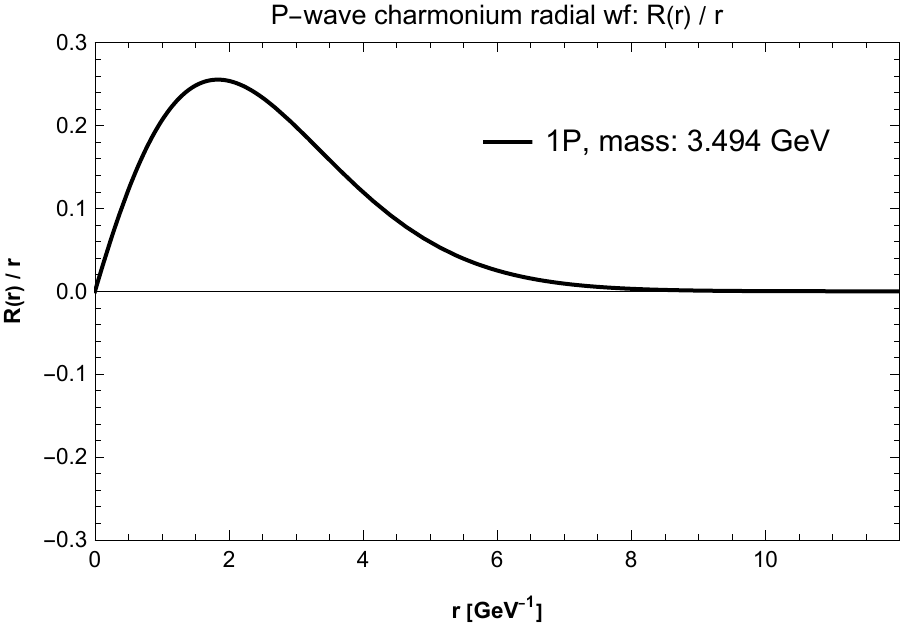}
\end{subfigure} 
\caption{S-wave and P-wave charmonium radial wavefunctions obtained according to  Sec.~\ref{subsec:Quarkonium}.}
\label{fig:ccbar}
\end{figure}

\begin{figure}[ht]
\begin{subfigure}{.4\textwidth}
  \centering
  \includegraphics[width=0.85\linewidth]{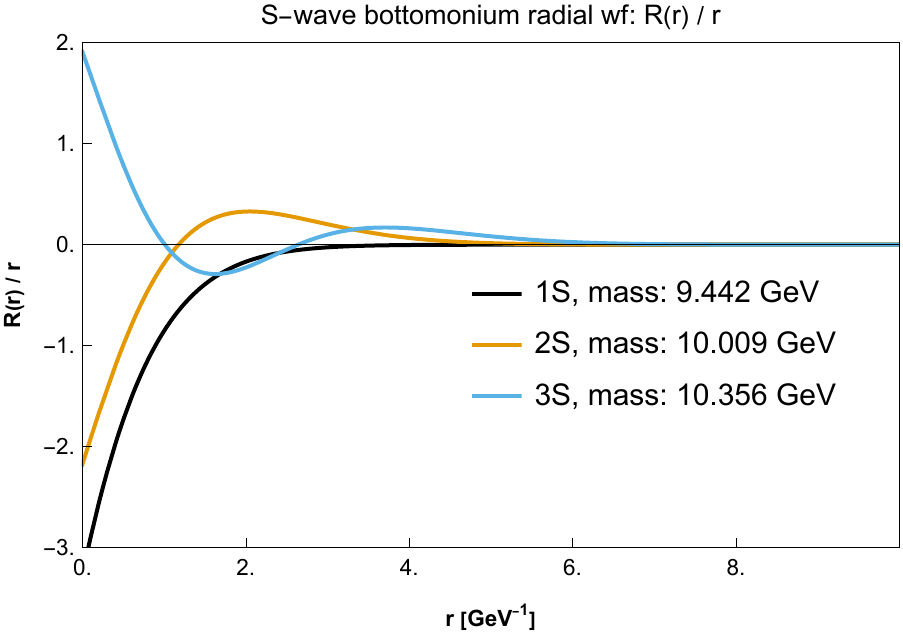}
\end{subfigure}
\begin{subfigure}{.4\textwidth}
  \centering
  \includegraphics[width=0.85\linewidth]{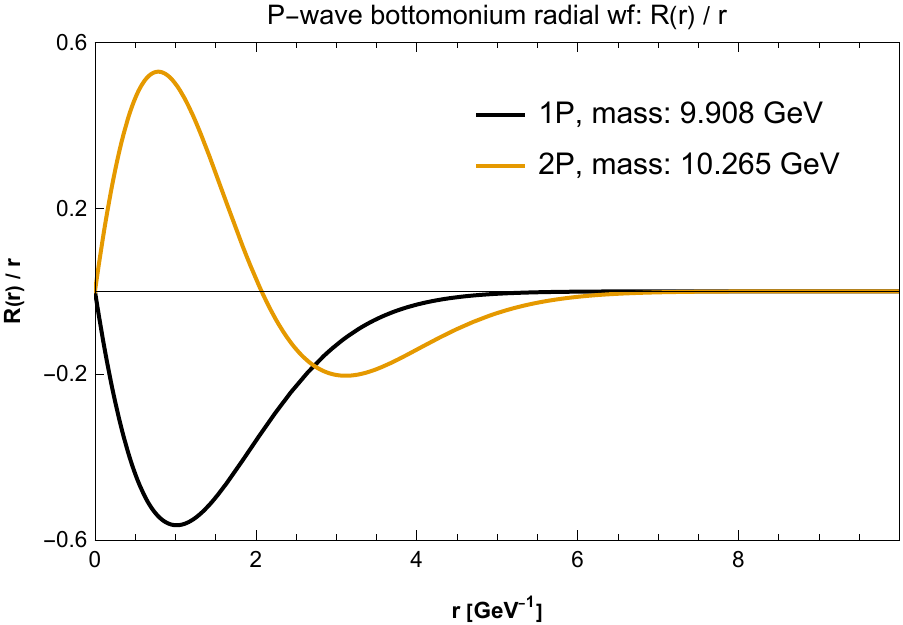}
\end{subfigure}
\caption{S-wave and P-wave bottomonium radial wavefunctions obtained according to Sec.~\ref{subsec:Quarkonium}.}
\label{fig:bbbar}
\end{figure}

We show in Figs.~\ref{fig:ccbar} and~\ref{fig:bbbar} 
charmonium and bottomonium S- and P-wave radial wavefunctions, obtained from the Schr\"odinger equation discussed in  Sec.~\ref{subsec:Quarkonium}.

\section{Hybrid wavefunctions}\label{app:hybridwf}
\begin{figure}[t]
\begin{subfigure}{.4\textwidth}
  \centering
  \includegraphics[width=0.90\linewidth]{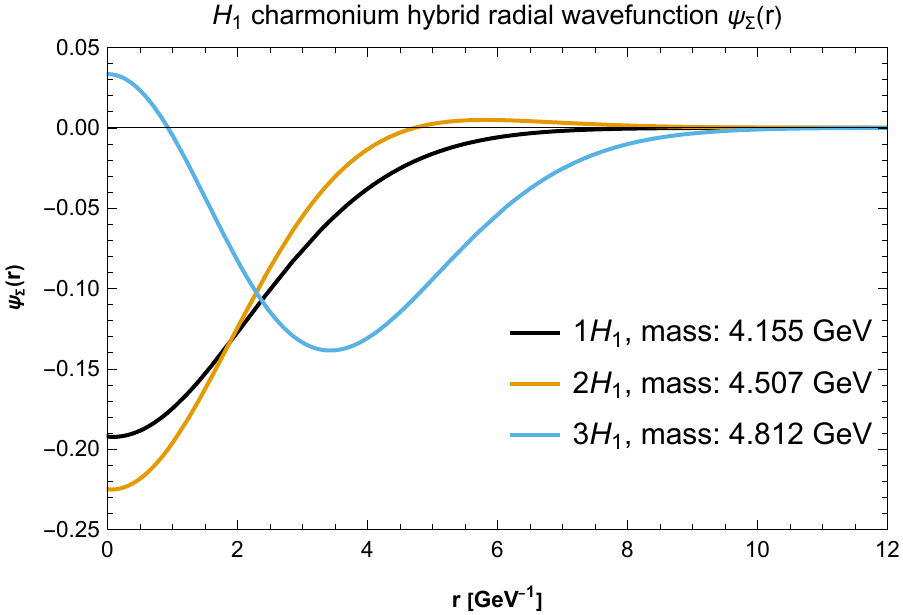}
\end{subfigure}
\begin{subfigure}{.4\textwidth}
  \centering
  \includegraphics[width=0.90\linewidth]{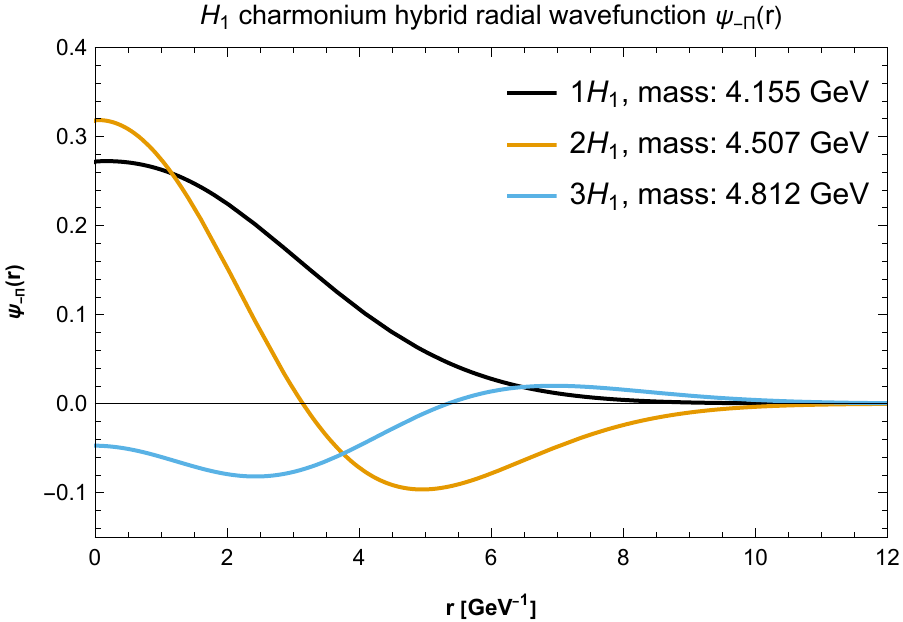}
\end{subfigure}
\caption{$H_1$ charmonium hybrid radial wavefunctions for $\Psi_{\Sigma}$ and $\Psi_{-\Pi}$. For details see Sec.~\ref{subsec:hybrids}.}
\label{fig:H1_charm}
\end{figure}

\begin{figure}[h!]
\begin{subfigure}{.4\textwidth}
  \centering
  \includegraphics[width=0.90\linewidth]{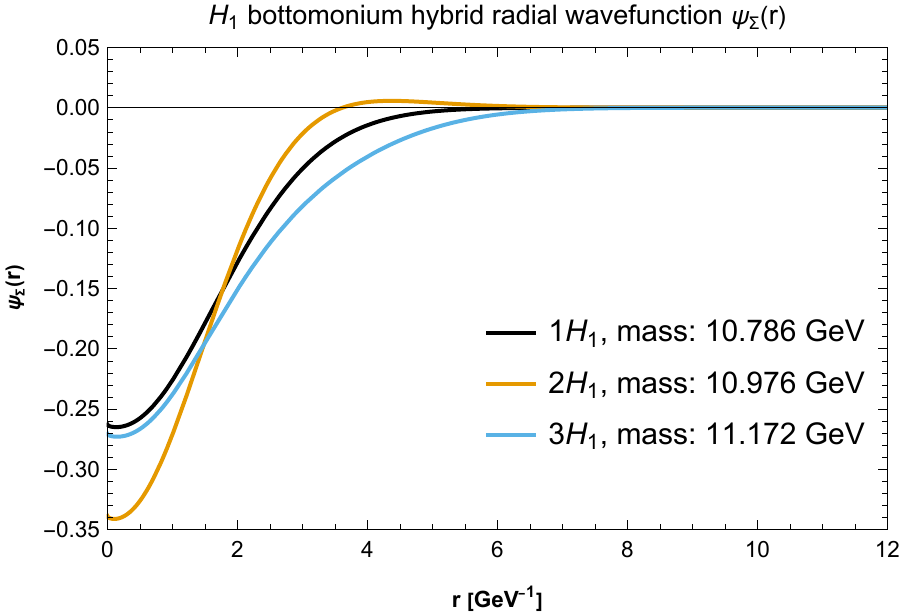}
\end{subfigure}
\begin{subfigure}{.4\textwidth}
  \centering
  \includegraphics[width=0.90\linewidth]{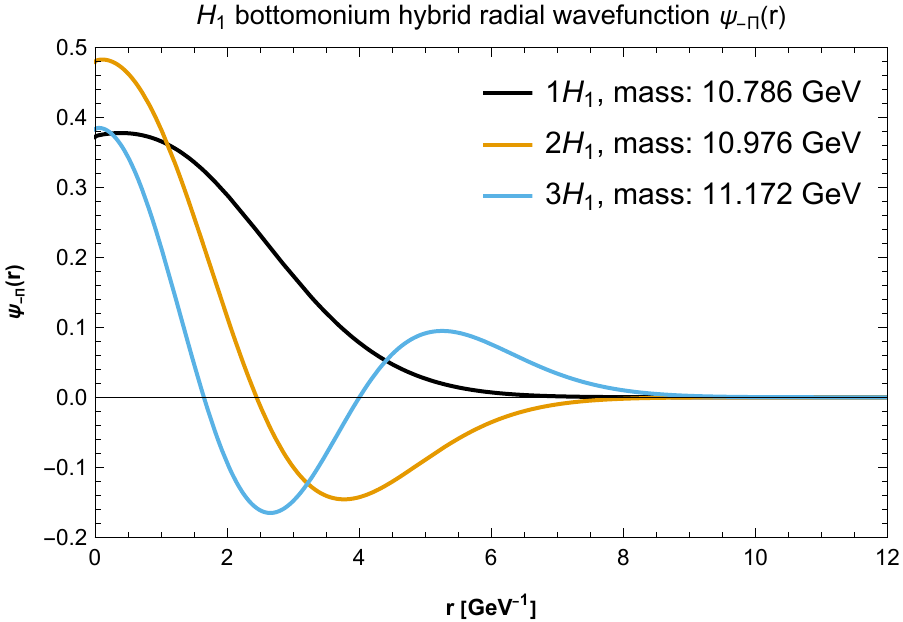}
\end{subfigure}
\caption{ $H_1$ bottomonium hybrid radial wavefunctions for $\Psi_{\Sigma}$ and $\Psi_{-\Pi}$. For details see Sec.~\ref{subsec:hybrids}.}
\label{fig:H1_bottom}
\end{figure}

\begin{figure}[h!]
\begin{subfigure}{.4\textwidth}
  \centering
\includegraphics[width=0.90\linewidth]{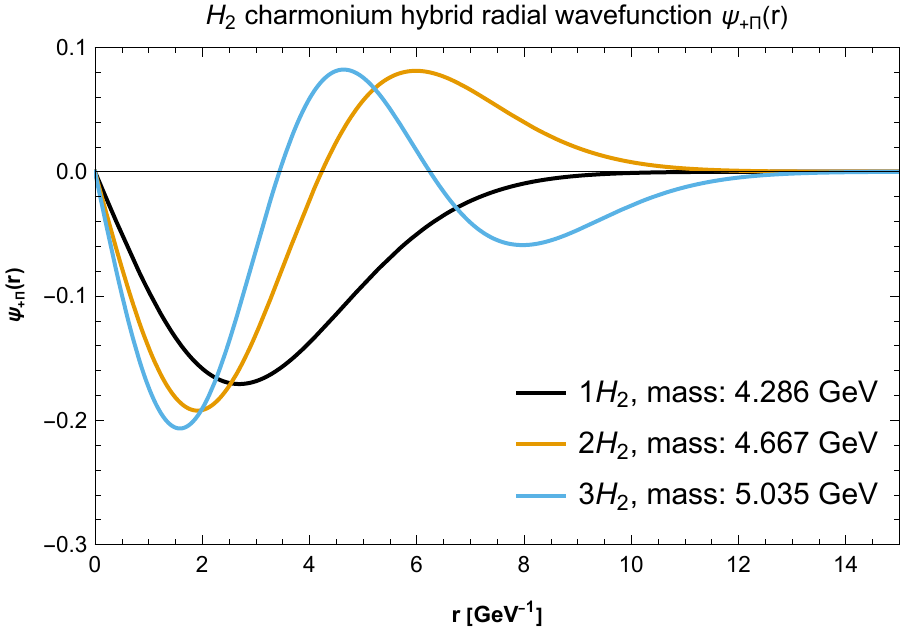}
\end{subfigure}
\begin{subfigure}{.4\textwidth}
  \centering
\includegraphics[width=0.90\linewidth]{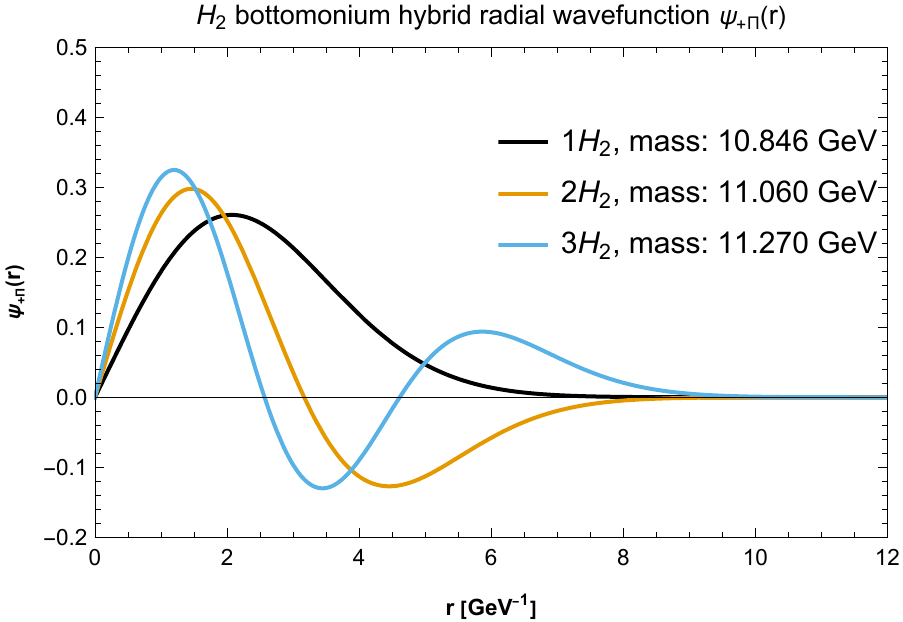}
\end{subfigure}
\caption{$H_2$ charmonium and bottomonium hybrid radial wavefunctions. 
For details see Sec.~\ref{subsec:hybrids}.}
\label{fig:H2-wave}
\end{figure}

\begin{figure}[ht]
\begin{subfigure}{.4\textwidth}
  \centering
\includegraphics[width=0.90\linewidth]{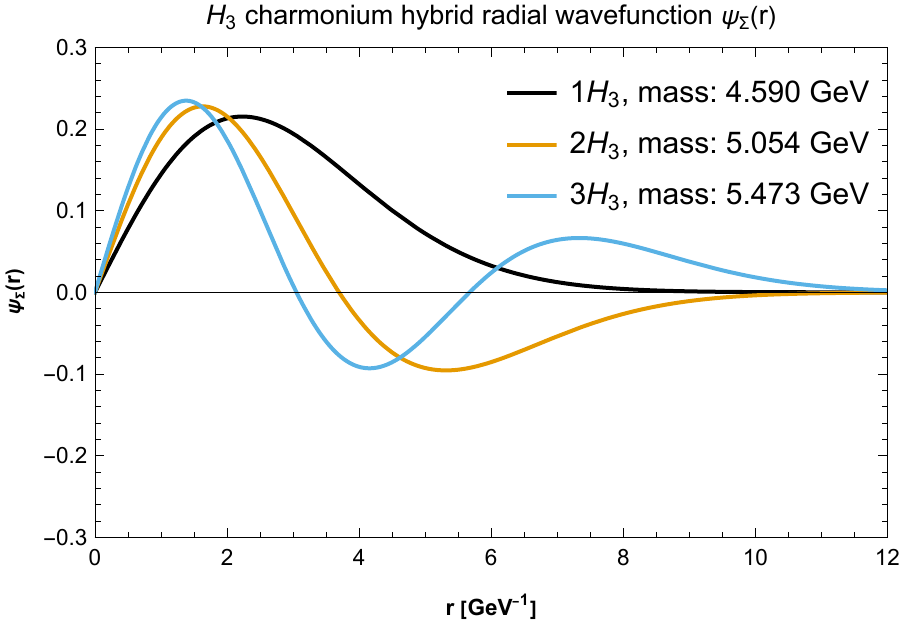}
\end{subfigure}
\begin{subfigure}{.4\textwidth}
  \centering
\includegraphics[width=0.90\linewidth]{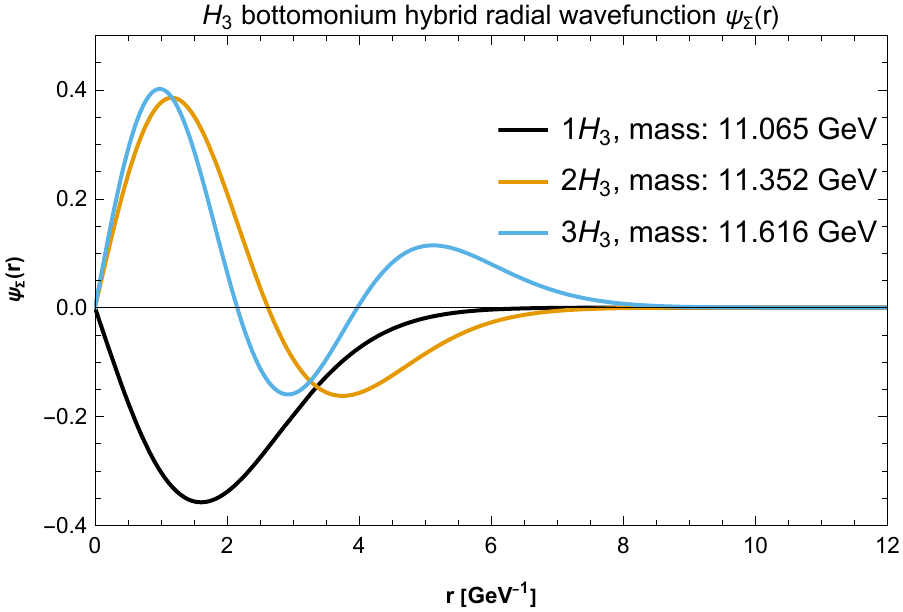}
\end{subfigure}
\caption{$H_3$ charmonium and bottomonium hybrid radial wavefunctions. 
For details see Sec.~\ref{subsec:hybrids}.}
\label{fig:H3-wave}
\end{figure}

\begin{figure}[ht]
\begin{subfigure}{.4\textwidth}
  \centering
  \includegraphics[width=0.90\linewidth]{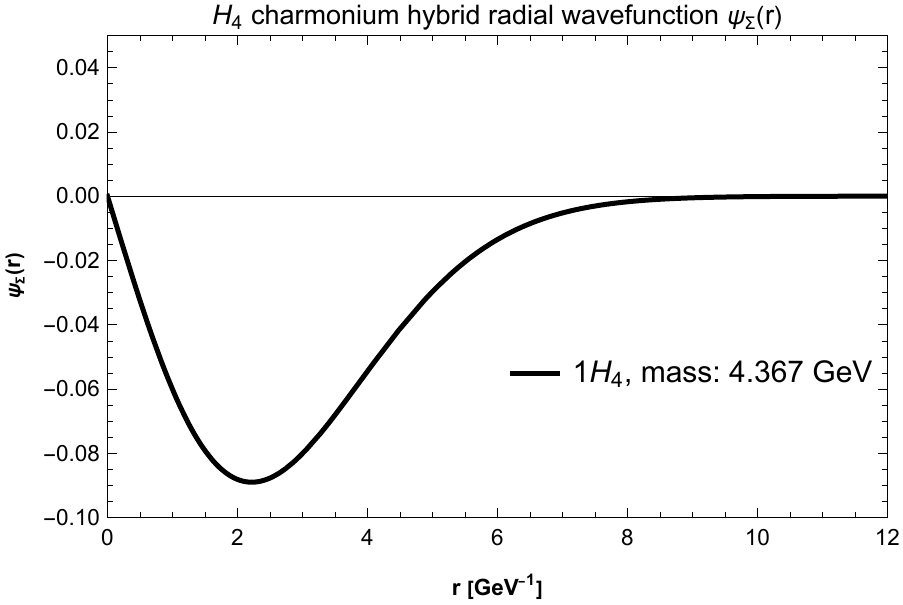}
\end{subfigure}
\begin{subfigure}{.4\textwidth}
  \centering
  \includegraphics[width=0.90\linewidth]{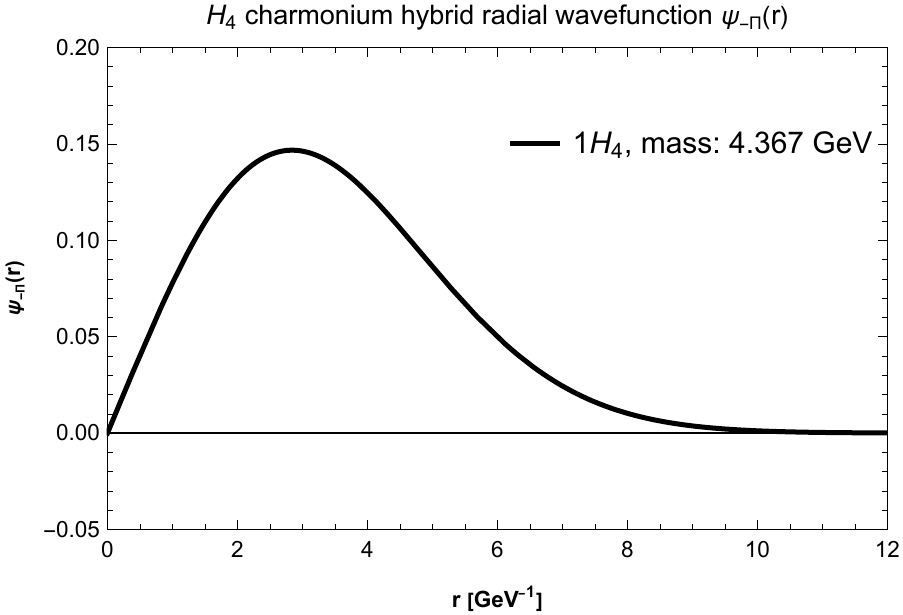}
\end{subfigure}
\caption{$H_4$ charmonium hybrid radial wavefunctions for $\Psi_{\Sigma}$ and $\Psi_{-\Pi}$. For details see Sec.~\ref{subsec:hybrids}.}
\label{fig:H4-wave}
\end{figure}

We show in Figs.~\ref{fig:H1_charm} and~\ref{fig:H1_bottom} 
the $H_1$ multiplet radial wavefunctions of charmonium and bottomonium hybrids, respectively. 
Similarly, in Fig.~\ref{fig:H2-wave} for the $H_2$ multiplet and in 
Fig.~\ref{fig:H3-wave} for the $H_3$ multiplet; 
in both figures, the left-hand side picture shows charmonium hybrid wavefunctions  
and the right-hand side picture shows bottomonium hybrid wavefunctions. 
Finally, in Fig.~\ref{fig:H4-wave} we show the $H_4$ multiplet radial wavefunctions of charmonium hybrids. 
The wavefunctions have been obtained according to the coupled Schr\"odinger equations discussed in Sec.~\ref{subsec:hybrids}.

\section{Gluonic correlator}\label{app:matching}
We approximate the correlator $\langle 0|G^{ia}(T/2)\phi^{ab}(T/2,t)E^{kb}(t) E^{lc}(t') \phi^{cd}(t',-T/2)G^{jd}(-T/2)|0\rangle$, appearing in Sec.~\ref{subsec:spin-conserving},
with the spectator gluon approximation that consists in neglecting the interaction between the low energy gluon fields $G^{ia}$ that constitute the hybrid and
the high energy gluon fields $E^{ia}$ that carry energy $\Delta E \gg \Lambda_{\rm QCD}$. 
This leads to factorize the correlator into a low energy 
two field correlator and into a high energy two field correlator computed 
in perturbation theory. At leading order and in the large time $T$ limit, we get
\begin{align}
&\langle 0|G^{ia}(T/2)\phi^{ab}(T/2,t)E^{kb}(t) E^{lc}(t') \phi^{cd}(t',-T/2)G^{jd}(-T/2)|0\rangle\nonumber\\
&\hspace{4cm} \approx 
\langle 0|G^{ia}(T/2)\phi^{ab}(T/2,-T/2) G^{jb}(-T/2)|0\rangle \, \frac{\delta^{kl}}{3} \int\frac{d^3k}{(2\pi)^3} |\bm{k}|e^{-i|\bm{k}|(t-t')}\nonumber\\
&\hspace{4cm} \approx
\delta^{ij}\frac{\delta^{kl}}{3} \, e^{-i\Lambda T} \int\frac{d^3k}{(2\pi)^3} |\bm{k}|e^{-i|\bm{k}|(t-t')}\,.
    \label{eq:gluon}
\end{align}

\section{Spin-matrix elements}\label{app:spin-matrix}
The spin part of the wavefunction is $\chi^{ab}=\xi^a\eta^b$, where the two-component spinors $\xi$ and $\eta$ transform as $\xi \to U \xi$, $\eta \to U^* \eta$ under SO(3), and $\chi$ transforms as $U\chi U^\dagger$.
The spin operators for $\xi$ and $\eta$ are $S_1^i=\sigma^i_1/2$ and $S_2^i=- \sigma^{iT}_2/2$, respectively.
For example, the expectation value of $S_1^iS_2^j$ is
\begin{equation}
  \xi^\dagger S_1^i \xi \eta^\dagger S_2^j \eta =-\frac{1}{4}\xi^{*a}(\sigma_1^i)^{ab}\xi^b\eta^{*c}(\sigma^{jT}_2)^{cd}\eta^d
                                                  =-\frac{1}{4}\xi^{*a}(\sigma^i_1)^{ab}\xi^b\eta^{d}(\sigma_2^j)^{dc}\eta^{*c}
                                                 =-\frac{1}{4}{\rm Tr}[\chi^\dagger\sigma^i_1\chi\sigma^j_2]\,.\label{eq:SS}
\end{equation}
We can calculate the spin part of $\langle Q_n|(S_1^i-S_2^i)|H_m\rangle$ as
\begin{align}
\langle Q_n|(S_1^i-S_2^i)|H_m\rangle&=\textrm{Tr}\left[\chi^\dagger_{Q_n}\frac{\sigma^i_1}{2}\chi_{H_m}\right]-
\textrm{Tr}\left[\chi^\dagger_{Q_n}\chi_{H_m}\left(-\frac{\sigma^i_2}{2}\right)\right].
\end{align}
Therefore, we have
\begin{align}
\langle S_{Q_n}=0|(S_1^i-S_2^i)|S_{H_m}=0\rangle&=0\,,\\
\langle S_{Q_n}=0|(S_1^i-S_2^i)|S_{H_m}=1\rangle&=\epsilon_H^i\,,\\
\langle S_{Q_n}=1|(S_1^i-S_2^i)|S_{H_m}=0\rangle&=\epsilon_Q^i\,,\\
\langle S_{Q_n}=1|(S_1^i-S_2^i)|S_{H_m}=1\rangle&=0\,,
\end{align}
where ${\bm \epsilon}_H$ and ${\bm \epsilon}_Q$ are the polarization vectors for spin-$1$ hybrid and quarkonium states.
The nonzero  averaged squared spin matrix elements are
\begin{align}
\frac{1}{2\cdot 1+1}\sum_{m_{S_{H_m}}}\langle S_{Q_n}=0|(S_1^i-S_2^i)|S_{H_m}=1,m_{S_{H_m}}\rangle \langle S_{H_m}=1,m_{S_{H_m}}|(S_1^i-S_2^i)|S_{Q_n}=0\rangle&=1\,,
\label{eq:<S>_Sh_1}\\
\frac{1}{2\cdot 0+1}\sum_{m_{S_{Q_n}}}\langle S_{Q_n}=1,m_{S_{Q_n}}|(S_1^i-S_2^i)|S_{H_m}=0\rangle \langle S_{H_m}=0|(S_1^i-S_2^i)|S_{Q_n}=1,m_{S_{Q_n}}\rangle&=3\,.
\label{eq:<S>_Sh_0}
\end{align}

\bibliography{Hybrid_decay}
\end{document}